\documentclass{template_SCOR_article}

\newcommand{\vero}[1]{{\color{black}#1}}
\newcommand{\verob}[1]{{\color{black}#1}}
\newcommand{\verop}[1]{{\color{black}#1}}

\def \ind{\mathds{1}}

\def \EE{\mathbb{E}}

\def \RR{\mathbb{R}}
\def \PP{\mathbb{P}}

\def \var{\mathrm{Var}}

\def \cD{\mathcal{D}}
\def \cE{\mathcal{E}}

\def \cJ{\mathcal{J}}
\def \cL{\mathcal{L}}

\def \cN{\mathcal{N}}

\def \cS{\mathcal{S}}

\def \bbeta{\bm \beta}

\def \bmu{\bm \mu}

\def \bSigma{\bm \Sigma}

\def \bX{\mathbf X}

\def \sfT{\mathsf{T}}

\title{On quantile oriented sensitivity analysis}
\date{\small \today}

\begin{document}

\maketitle

\begin{abstract}%   <- trailing '%' for backward compatibility of .sty file
We propose to study quantile oriented sensitivity indices (QOSA indices) and quantile oriented Shapley effects (QOSE). Some theoretical properties of QOSA indices will be given and several calculations of QOSA indices and QOSE will allow to better understand the behaviour and the interest of these indices. 
\end{abstract}

\underline{Keywords:} 
Quantile Oriented Sensisitivy Analysis, Shapley effects. \\
\ \\

\section{Introduction}

Sensitivity Analysis (SA) is defined by \citet{saltelli2004sensitivity} as ``the study of how the uncertainty in the output of a model can be apportioned to different sources of uncertainty in the model input''. Various tools exist today to perform a SA (see e.g. \citet{iooss2015review} for a review of SA methods).
We are especially interested in \textit{Global Sensitivity Analysis methods} - GSA -  which allow to study the effects of simultaneous variation of the inputs on the model output in their entire domain. 
%Thus, one can look closer at both output variations induced by individual inputs and/or interactions between several of them (i.e., groups of input variables). We are specifically interested in Global Sensitivity Analysis. 
For a detailed description of  sensitivity analysis methods, the interested reader can refer to the various survey papers dedicated to this topic \citep{saltelli2004sensitivity,saltelli2008global,faivre2016analyse,borgonovo2016sensitivity,borgonovo2017sensitivity}. Variance-based methods are common tools in the analysis of complex physical phenomenons. Most of them rest on an ANalysis Of VAriance (ANOVA) of the model output and give information on the sensitivity around the mean (as it is variance based). In this paper, we are interested in Quantile Oriented indices, in order to obtain informations on the sensitivity around quantiles. Much less work has been done on Quantile Oriented Sensitivity Analysis (QOSA). We shall focus on indices defined firstly in \citet{fort2016new}.  \\
\ \\
For completeness, \Cref{sec:3:variance_based_methods} shortly recalls variance-based methods dealing both with indepedent and dependent inputs.   \Cref{sec:5:qosa_indices} introduces QOSA indices which allows to quantify the sensitivity over a quantile. Note that in \citet{kucherenko2019quantile}, other quantile oriented indices have been defined. Nevertheless, QOSA indices have more natural interpretations as will be seen in \Cref{sec:5:qosa_indices}. Some properties of QOSA indices are also proposed within this section. Several calculations of QOSA indices are done in \Cref{sec:qosa_calculs} and a preliminary work is carried out in order to understand the impact of the statistical dependence between the inputs over these indices. Facing some interpretation issues, we finally propose in \Cref{sec:6:shapley_qosa} new indices based on Shapley values \citep{shapley1953value} which seem to be a promising alternative. \Cref{sec:conclusion} presents some further perspective of research. \\
\ \\
Consider a model $Y = \eta(\bX)$ with $d$ random inputs denoted by $ \bX = \left( X_1, X_2, \dots, X_d \right)$. Let $\bX_{\cJ}$ indicate the vector of inputs corresponding to the index set $\cJ \subseteq \cD$ where $\cD = \{ 1, 2, \dots, d \}$. GSA aims at quantifying the impact of the inputs $ X_1, X_2, \dots, X_d $ on the output $Y$. 

\section{Variance-based sensitivity indices}\label{sec:3:variance_based_methods}

We shall briefly recall the framework of Sobol indices and Shapley effects. 
\subsection{Sobol indices}\label{subsec:3:sobol_indices_idt}

Sobol' sensitivity indices stem from the works of \citet{fisher1923studies} and \citet{Hoeffding48} on the U-statistics taken up by various authors over time such as \citet{efron1981jackknife}. Those ultimately lead to a functional ANOVA expansion of the model output $\eta$:
\begin{equation}\label{eq:3:HDMR}
\eta (\bX) = \eta_0 + \sum_{i=1}^d \eta_i(X_i) + \sum_{1 \leqslant i < j \leqslant d} \eta_{ij} (X_i, X_j) + \cdots + \eta_{1, \dots, d}(\bX) \ .
\end{equation}
This decomposition is unique under orthogonality constraints (see \citet{sobol1993sensitivity}), \vero{that are verified when dealing with independent inputs}, which lead to the decomposition of the global variance as
\begin{equation}\label{eq:3:ANOVA}
\var(Y) = \sum_{i=1}^d V_i + \sum_{1 \leqslant i < j \leqslant d} V_{ij} + \dots + V_{1, \dots, d} \ ,
\end{equation}
where
\begin{align*}
V_i &= \var \left( \eta_i \left( X_i \right) \right) = \var \left( \EE \left[ \left. Y \right| X_i \right] \right) \ , \\
V_{ij} &= \var \left( \eta_{ij} \left( X_i, X_j \right) \right) = \var \left( \EE \left[ \left. Y \right| X_i, X_j \right] \right) - V_i - V_j \ , \\
V_{ijk} &= \var \left( \eta_{ijk} \left( X_i, X_j, X_k \right) \right) = \var \left( \EE \left[ \left. Y \right| X_i, X_j, X_k \right] \right) - V_{ij} - V_{ik} - V_{jk} - V_i - V_j - V_k \ , \\
& \vdots \\
V_{1, \dots, d} &= \var \left( \eta_{1, \dots, d}(\bX) \right) = \var \left( Y \right) - \sum_{i=1}^d V_i - \sum_{1 \leqslant i < j \leqslant d} V_{ij} - \ldots - \sum_{1 \leqslant i_1 < \ldots < i_{d-1} \leqslant d} V_{i_1 \ldots i_{d-1}} \ .
\end{align*}
The so-called Sobol indices given in \citet{sobol1993sensitivity} are derived from \eqref{eq:3:ANOVA}:
\begin{equation}\label{eq:3:sobol_indices}
S_i = \dfrac{V_i}{\var(Y)}, \qquad S_{ij} = \dfrac{V_{ij}}{\var(Y)} , \qquad \dots
\end{equation}
where the first-order index $S_i$ (also called main effect) measures the part of variance of the model output that stems from the variability in $X_i$, the second-order index $S_{ij}$ measures the part of variance of the model output due to the interaction between $X_i$ and $X_j$ and so on for higher interaction orders.

Using the $S_i, S_{ij}$ and higher order indices given above, one can build a picture of the importance of each variable in the output variance. However, when the number of variables is large, this requires the evaluation of $2^d - 1$ indices, which can be too computationally demanding and whose interpretation becomes difficult. For this reason, another popular variance based coefficient called Total-order index by \citet{homma1996importance} is used.  It measures the contribution to the output variance of $X_i$, including its main effect as well as all its interaction effects, of any order, with any other input variables. This index is defined by
\begin{equation}\label{eq:3:total_sobol}
\begin{split}
ST_i &= S_i + \sum_{1 \leqslant i < j \leqslant d} S_{ij} + \dots + S_{1, \dots, d} \\
&= 1 - \dfrac{\var_{\bX_{-i}} \left( \EE_{X_i} \left[ \left. Y \right| \bX_{-i} \right] \right)}{\var(Y)} = \dfrac{\EE_{\bX_{-i}} \left[ \var_{X_i} \left( \left. Y \right| \bX_{-i} \right) \right]}{\var(Y)} \ ,
\end{split}
\end{equation}
where the notation $\bX_{-i}$ indicates the set of all variables except $X_i$. Note that the following property can easily be deduced : 
\[
0 \leqslant S_i \leqslant ST_i \leqslant 1 \ .
\]
Hence, the closer the index $ST_i$ is to 1, the more influential the variable is. It should be noted that the case of equality $S_i=ST_i$ occurs if we have a purely additive model.

Sobol indices are well-defined for independent inputs. Indeed, the functional decomposition of the output variance given in \eqref{eq:3:ANOVA} is unique in this context. This makes possible to clearly identify the contribution of each input or group of inputs to the variance output.

However, in many applications, it is common for inputs to have a statistical dependence structure imposed by a probabilistic dependence \verob{structure}. In such a case, the uniqueness of the functional ANOVA decomposition  is no longer guaranteed. The classical Sobol indices can still be calculated but their interpretation becomes difficult. Indeed, as mentioned in \citet{song2016shapley}, the sum of first-order effects may exceed the total variance of the output or the sum of the total effects may be lower than the total variance of the output.

Several works have been carried out to overcome this limitation and extend Sobol indices to the case of stochastic dependence such as \citet{caniou2012global,kucherenko2012estimation,mara2012variance,mara2015non}. However, none of these works has given an univocal definition of the functional ANOVA decomposition for dependent inputs as the one provided by \citet{sobol1993sensitivity} when inputs are independent. A new variable importance measure has been defined in \citet{chastaing2012generalized} through a generalization of ANOVA when inputs are dependent \citep{stone1994use}. But this measure comes with two conceptual problems: it requires some restrictive conditions on the joint probability distribution of the inputs as underlined in \citet{owen2017shapley} and its interpretation remains difficult because it can be negative.
New indices called Shapley effects have been proposed in \citet{owen2014sobol}. They present good properties in the presence of dependence: they are non negative, they sum to the total output variance and they are easy to interpret as highlighted by \citet{song2016shapley,iooss2019shapley}.

\subsection{Shapley effects}\label{subsec:3:shapley_effects}

\textit{Shapley values} have been introduced in game theory by \citet{shapley1953value}. The motivation, in the context of cooperative game theory, was to define an attribution method to allocate fairly the value created by a team effort to its individual members. Turning now to variance-based sensitivity analysis, it appears that the idea of assigning a portion of the output variance to each input variable has some similarities. These were highlighted and brought to the SA community, in the context of variance-based sensitivity analysis, by \citet{owen2014sobol}.

Formally, in \cite{song2016shapley} a $d$-player game with the set of players $\cD = \{ 1, 2, \dots, d \}$ is defined as a real-valued function that maps a subset of $\cD$ to its corresponding cost, i.e., $c:2^{\cD} \mapsto \RR$ with $c(\emptyset) = 0$. Hence, $c(\cJ)$ represents the cost that arises when the players in the subset $\cJ$ of $\cD$ participate in the game. Let $v^i = v^i(c), \ i = 1, \ldots, d$, be the Shapley value for each player that will be defined below. According to \cite{winter2002shapley}, an attribution method should have the next four compelling properties:
\begin{itemize}
	\item \textbf{Efficiency:} $\displaystyle\sum_{i=1}^d v^i = c \left( \cD \right)$. The sum of the values attributed to the players must be equal to what the coalition of all the players can obtain.
	\item \textbf{Symmetry:} if $c(\cJ \cup i) = c(\cJ \cup j)$ for all $\cJ \subseteq \cD \backslash \left\lbrace i, j \right\rbrace$, then $v^i = v^j$. The contribution of two players should be the same if they contribute equally to all possible coalitions.
	\item \textbf{Dummy:} if $c(\cJ \cup i) = c(\cJ)$ for all $\cJ \subseteq \cD \backslash \left\lbrace i \right\rbrace$, then $v^i=0$. A player who does not change the predicted value, no matter to which coalition of players it is added, should have a contribution value of 0.
	\item \textbf{Additivity:} if the $i$-th player has a contribution $v^i$ (resp. $v'^i$) in the coalitional game described by the gain function $c$ (resp. $c'$). Then, the contribution of the $i$-th player in the new coalitional game described by the gain function $c+c'$ is $v^i + v'^i$ for $i \in \cD$.
\end{itemize}

\cite{shapley1953value} showed that the unique valuation $v^i$ that satisfies these properties is \verob{given by}
\begin{equation}\label{eq:3:shapley_value}
v^{i} = \sum_{\cJ \subseteq \cD \backslash \{i\}} \dfrac{(d - |\cJ| - 1)!|\cJ|!}{d!} \left( c \left( \cJ \cup \{i\} \right) - c \left( \cJ \right) \right) \ ,
\end{equation}
defined for the player $i$ with respect to \verob{the cost function} $c(\cdot)$ and where $|\cJ|$ indicates the size of $\cJ$. In other words, $v^{i}$ is the incremental cost of including player $i$ in \verob{the set $\cJ$ averaged over all sets $\cJ \subseteq \cD \backslash \{i\}$. } %\verop{It should also be noted that the weight for each incremental cost of size-$s$ subset of $\cD \backslash \{i\}$ in \eqref{eq:3:shapley_value} can be written as $\frac{(d - s - 1)!s!}{d!} = \frac{1}{d} \binom{d-1}{s}^{-1}$}.

In the framework of global sensitivity analysis, we \verob{may} consider the set of inputs of $\eta(\cdot)$ as the set of players $\cD$. We then need to define a $c(\cdot)$ cost function such that for $\cJ \subseteq \cD$, $c(\cJ)$ measures the part of variance of $Y$ caused by the uncertainty of the inputs in $\cJ$. To this aim, we want a cost function that verifies $c(\emptyset) = 0$ and $c(\cD) = \var (Y)$.

\cite{owen2014sobol}  proposed the cost function $\tilde{c}(\cJ) = \var \left( \EE \left[ Y | \bX_{\cJ} \right] \right)$ for considering the Shapley value in the framework of variance-based sensitivity indices. \cite{song2016shapley}  showed in their Theorem 1 that the Shapley values defined using cost functions $\tilde{c}(\cJ)$ and $c(\cJ) = \EE \left[ \var \left( Y | \bX_{-\cJ} \right) \right]$ are equal. They used the term \textit{Shapley effects} to describe variance based Shapley values as new importance measures in SA. Note that in \cite{owen2014sobol,song2016shapley}, the cost function is not normalized by the variance of $Y$, whereas, in this paper, we consider its normalized version to quantify relative importance of each input with respect to the output variance. We denote hereafter the \textit{Shapley effect} by $Sh^i$ and a generic \textit{Shapley value} by $v^i$. \\
The purpose of  Sobol indices defined in Subsection \ref{subsec:3:sobol_indices_idt} is to decompose $\var(Y)$ and allocate it to \textit{each subset} $\cJ \subseteq \cD$ whereas Shapley effects decompose $\var(Y)$ and allocate it to \textit{each input} $X_i$. This difference allows to consider any variables regardless of their dependence with other inputs.  Shapley effects rely on an equitable allocation of part of the output variance to each input. %During this allocation process, interaction and dependence effects of a subset of inputs are fairly shared with each individual input within the subset. This fair allocation results in Shapley effects being non negative and sum-up to one, allowing an easy interpretation for ranking input factors. Each one can therefore be interpreted as a measure of the part of the variance of $Y$ related to the $i$-th input of $\eta$. Hence, the closer the index $Sh^i$ is to 1, the more influential the variable is.
%It has to be noted that several studies have been carried out on these new indices. In the first place, in case of independent inputs, \cite{owen2014sobol} showed the Shapley effects are bounded by the Sobol indices, the first-order (resp. total) index as lower (resp. upper) bound. In the second place,  
In case of dependent inputs, several test-cases where the Shapley effects can be analytically computed have been investigated in \cite{owen2017shapley,iooss2019shapley,benoumechiara2019shapley} in order to understand the effect of the dependence between inputs on the variance based Shapley values. These studies highlighted several properties of these indices. \cite{benoumechiara2019shapley} have also compared the Shapley effects with the strategy proposed in \cite{mara2015non} based on the estimation of four sensitivity indices per input. In this last case, the Shapley effects can be a good alternative to the existing extensions of classical Sobol indices. Indeed, Shapley effects allow an apportionment of the interaction and dependence contributions between the input involved, making them condensed and easy-to-interpret indices. At last, we can mention that \citet{rabitti2019shapley} have recently introduced Shapley-Owen interaction effects which are a generalization of Shapley effects in order to study the synergistic/antagonistic nature of interactions among inputs.

\section{QOSA indices}\label{sec:5:qosa_indices}

In the previous section, we have reviewed variance-based measures. Even if these indices are extremely popular and informative importance measures, they only study the impact of the input variables around the expectation of the output distribution as they use  the \verob{variance as a distance measure}. However, as highlighted by \citet{borgonovo2006measuring}, in some cases, this one poorly represents the variability/uncertainty of the output distribution. Different approaches have been developped to overcome this issue including, for example, moment independent importance measures proposed by \citet{borgonovo2007new,borgonovo2011moment} that quantify the influence of an input over the whole distribution of the output. 

Another approach presented in \citet{fort2016new} is to define indices that quantify the impact of inputs \verob{$\bX = \left( X_1, \ldots, X_d \right)$ on a feature} of interest of the output distribution depending on the problem (mean, quantiles and so on). They refer to this method as \textit{Goal-Oriented Sensitivity Analysis} (GOSA). These indices rely on \verob{contrast functions} and are members of a wider family containing sensitivity indices based on dissimilarity measures \citep{da2015global}. We shall consider the specific case of  \textit{Quantile Oriented Sensitivity Analysis} (QOSA).

\subsection{First-order QOSA indices}\label{sub:sec:qosa_def}
Let us focus on QOSA indices measuring the impact of the inputs over the $\alpha$-quantile of the output distribution. Given a level of quantile $\alpha \in \left] 0, 1 \right[$, let us recall the expression of the first-order QOSA index:
\begin{equation}\label{eq:5:qosa_index}
\begin{split}
S_i^\alpha &= \dfrac{\displaystyle \min_{\theta \in \RR} \EE \left[ \psi_\alpha \left( Y,\theta \right) \right] - \EE \left[ \min_{\theta \in \RR} \EE \left[ \left. \psi_\alpha \left( Y,\theta \right)\right| X_i \right] \right]}{\displaystyle \min_{\theta \in \RR} \EE \left[ \psi_\alpha \left(Y,\theta \right) \right]} \\
&= \dfrac{\EE \left[ \psi_\alpha \left( Y, q^\alpha\left( Y \right) \right) \right]-\EE \left[ \psi_\alpha \left( Y, q^\alpha \left( \left. Y \right| X_i \right) \right) \right]}{\EE \left[ \psi_\alpha \left( Y, q^\alpha\left( Y \right) \right) \right]} \ ,
\end{split}
\end{equation}
where $\psi_\alpha: (y,\theta) \mapsto (y-\theta) \left(\alpha-\ind_{\left\lbrace y \leqslant \theta \right\rbrace} \right)$ is the contrast function associated to the $\alpha$-quantile and  the $q$'s are the quantiles
\[
q^\alpha \left( Y \right) = \argmin_{\theta \in \RR} \EE \left[ \psi_\alpha \left(Y,\theta \right) \right] \quad \textnormal{ and } \quad q^\alpha \left( \left. Y \right| X_i=x_i \right) = \argmin_{\theta \in \RR} \EE \left[ \left. \psi_\alpha \left( Y,\theta \right) \right| X_i=x_i \right] \ .
\]
Remark that replacing $\psi_\alpha$ in the above equation \verob{by $(y\/,\theta) \mapsto (y-\theta)^2$} leads to the definition of first-order Sobol indices. In order to \vero{interpret} QOSA indices, one has to consider $\psi_\alpha(Y,\theta)$ as a dispersion measure of $Y$ which is minimized for $\theta=q^\alpha(Y)$. So that QOSA indices compare the dispersion of $Y$ around its quantile with its conditional counterpart. \\  
The first-order QOSA \vero{indices} have been defined in \citet{fort2016new}, studied and estimated in \citet{browne2017estimate,maume2018estimation,kedc_estimation}. They may be \vero{rewritten} as follows
\[
S_i^\alpha = \dfrac{\EE \left[ Y \ind_{\left\lbrace Y \leqslant q^\alpha \left( \left. Y \right| X_i \right) \right\rbrace} \right] - \EE \left[ Y \ind_{\left\lbrace Y \leqslant q^\alpha \left( Y \right) \right\rbrace} \right]}{\alpha \EE \left[ Y \right] - \EE \left[ Y \ind_{\left\lbrace Y \leqslant q^\alpha \left( Y \right) \right\rbrace} \right]} = 1-\dfrac{\alpha \EE \left[ Y \right] - \EE \left[ Y \ind_{\left\lbrace Y \leqslant q^\alpha \left( \left. Y \right| X_i \right) \right\rbrace} \right]}{\alpha \EE \left[ Y \right] - \EE \left[ Y \ind_{\left\lbrace Y \leqslant q^\alpha \left( Y \right) \right\rbrace } \right]} \ .
\]

Let us mention that \citet{kucherenko2019quantile} have recently proposed new indices to assess the impact of inputs over the $\alpha$-quantile of the output distribution. Instead of considering the expression of the first-order Sobol index based on a contrast function as in \eqref{eq:5:qosa_index}, they consider the expression of Sobol' indices with numerator  $\var \left( \EE \left[ \left. Y \right| X_i \right] \right) = \EE \left[ \left( \EE \left[ \left. Y \right| X_i \right] - \EE \left[ Y \right] \right)^2 \right]$ and simply replace the expectations by $\alpha$-quantiles to define the following indices
\[
\bar{q}_{i,1}^\alpha = \EE \left[ \left| q^\alpha \left( Y \right) - q^\alpha \left( \left. Y \right| X_i \right) \right| \right] \quad \textnormal{and} \quad
\bar{q}_{i,2}^\alpha = \EE \left[ \left( q^\alpha \left( Y \right) - q^\alpha \left( \left. Y \right| X_i \right) \right)^2 \right] \ .
\]
They also provide the normalized versions as follows
\[
Q_{i,1}^\alpha = \dfrac{\bar{q}_{i,1}^\alpha}{\sum\limits_{j=1}^d \bar{q}_{j,1}^\alpha} \quad \textnormal{and} \quad
Q_{i,2}^\alpha = \dfrac{\bar{q}_{i,2}^\alpha}{\sum\limits_{j=1}^d \bar{q}_{j,2}^\alpha} \ .
\]
These measures thereby quantify the mean distance between quantiles $q^\alpha \left( Y \right)$ and $q^\alpha \left( \left. Y \right| X_i \right)$ rather than the mean distance between average contrast functions like in the first-order QOSA index given in \eqref{eq:5:qosa_index}. The indices $\bar{q}_{i,1}^\alpha$ and $Q_{i,1}^\alpha$ will be called below {\em absolute value indices};  $\bar{q}_{i,2}^\alpha$ and $Q_{i,2}^\alpha$ will be called below {\em squared indices}. The example below shows that their interpretation as sensitivity indices is questionnable, it is why we shall focus on QOSA indices. \\
\ \\
Consider the simple model also studied in \citet{fort2016new}: $Y=X_1-X_2$ where $X_1$ and $X_2$ are two independent exponential random variables with expectation $1$. So that $Y$ follows a Laplace distribution. QOSA indices have closed form formulas that may be found in \citet{fort2016new}. The indices proposed in \citet{kucherenko2019quantile} may also be computed. \vero{Indeed, with 
\[
\gamma_1 =
\begin{cases}
-\log \left( 2 \alpha \left( 1-\alpha \right) \right) \textnormal{ if } \alpha\geq \frac12 \\
\log(2) \textnormal{ if } \alpha<\frac12
\end{cases}
,
\]
and 
\[
\gamma_2 =
\begin{cases}
\log(2) \textnormal{ if } \alpha \geq \frac12 \\
-\log \left( 2 \alpha \left( 1-\alpha \right) \right) \textnormal{ if } \alpha < \frac12
\end{cases}
,
\]
we have
\[
\bar{q}_{1,1}^\alpha = \gamma_1+2e^{-\gamma_1}-1 \quad \textnormal{and} \quad
\bar{q}_{2,1}^\alpha= \gamma_2+2e^{-\gamma_2}-1 \ ;
\]
\[
\bar{q}_{1,2}^\alpha = \gamma_1^2 - 2\gamma_1 + 2 \quad \textnormal{and} \quad
\bar{q}_{2,2}^\alpha = \gamma_2^2 - 2 \gamma_2 + 2 \ .
\]
}
Below we show the behaviour of QOSA, $\bar{q}_{i,1}^\alpha$ and $\bar{q}_{i,2}^\alpha$ indices. As expected, QOSA indices show that $X_1$ has more influence on quantiles of level higher that $\frac12$ and $X_2$ is more influent for $\alpha$ \vero{lower than} $\frac12$. The interpretation of $\bar{q}_{i,j}^\alpha$ is not so clear since $\bar{q}_{1,j}^\alpha$ are constant for $\alpha$ less than $\frac12$ and $\bar{q}_{2,j}^\alpha$ are constant for $\alpha$ greater than $\frac12$. \vero{Moreover}, \verob{the $\bar{q}_{i,2}^\alpha$'s are not monotonic}. \\
\begin{center}
\includegraphics[scale=0.3]{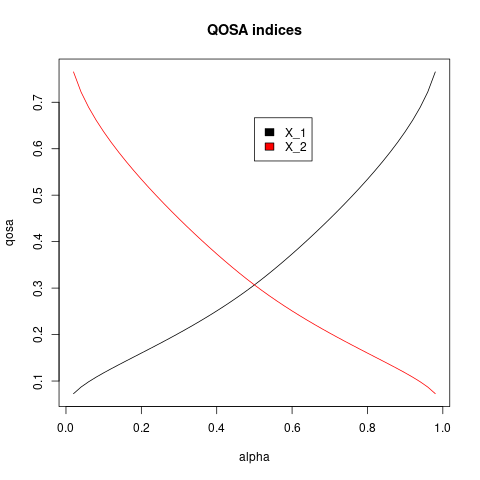} \ \includegraphics[scale=0.3]{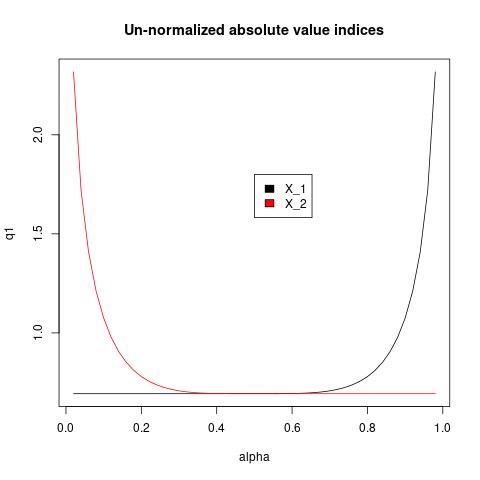} \ \includegraphics[scale=0.3]{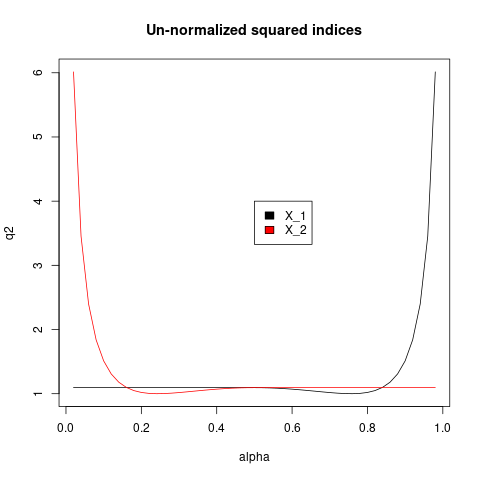}
\end{center}
The normalized indices below show that the interpretation for $Q_{i,j}^\alpha $ \verop{remains} questionnable while the normalized QOSA indices keep the interpretation of the un-normalized ones. 
\begin{center}
\includegraphics[scale=0.3]{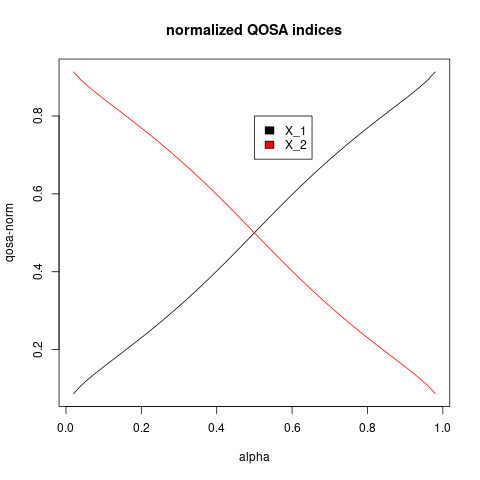} \ \includegraphics[scale=0.3]{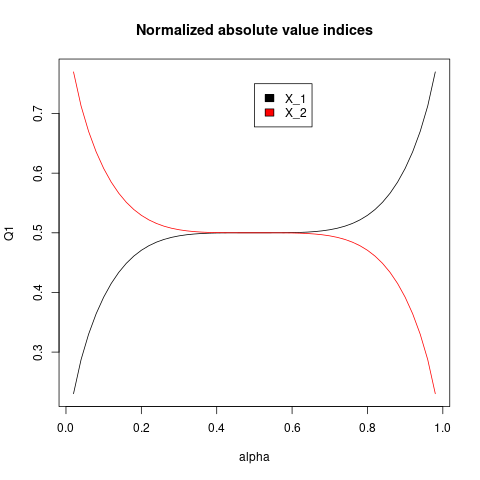} \ \includegraphics[scale=0.3]{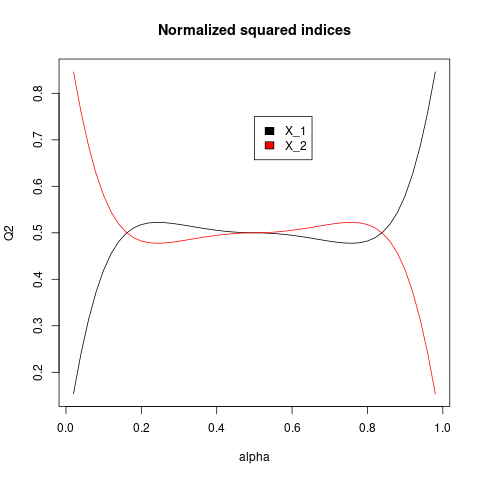}
\end{center}

\subsection{Some elementary properties of QOSA indices}\label{sub:sec:qosa_ppte}
Let us turn to some elementary properties of QOSA indices. \\
The following lemma is useful, it is closely related to the proof of sub-additivity of TVaR in risk theory \verob{(see \citet{marceau2013modelisation} e.g.)}.
\begin{lem}\label{lem:5:like_TVaR}
Consider any event $E$ such that $\PP(E)=\alpha$. Then, for any random variable $X$, we have
\[
\EE \left[ X \ind_{\left\lbrace X \leqslant q^\alpha(X)\right\rbrace} \right] \leqslant \EE \left[ X \ind_E \right] \ ,
\]
with $q^\alpha(X)$ the $\alpha$-quantile of $X$.
\end{lem}
\begin{myproof}[Proof of \Cref{lem:5:like_TVaR}]
We have:
\begin{align*}
\EE \left[ X \ind_{\left\lbrace X \leqslant q^\alpha(X) \right\rbrace} \right] - \EE \left[ X\ind_E \right] & = \EE \left[ X \left( \ind_{\left\lbrace X \leqslant q^\alpha(X)\right\rbrace} - \ind_E \right) \right] \\
&= \EE \left[ \left( X - q^\alpha(X) \right) \left( \ind_{\left\lbrace X\leqslant q^\alpha(X)\right\rbrace} - \ind_E \right) \right] \leqslant 0 \ .
\end{align*}
\end{myproof}

As a consequence, since $\PP \left( \left. Y \leqslant q^\alpha \left( \left. Y \right| X_i \right) \right| X_i \right) = \alpha$ and $\PP \left( Y \leqslant q^\alpha \left( Y \right) \right) = \alpha$, we get that
\begin{align}
& \alpha \EE \left[ Y \right] - \EE \left[ Y \ind_{\left\lbrace Y \leqslant q^\alpha \left( \left. Y \right| X_i \right) \right\rbrace } \right] = \EE \left[ \left( Y-q^\alpha \left( \left. Y \right| X_i \right) \right) \left( \alpha - \ind_{\left\lbrace Y \leqslant q^\alpha \left( \left. Y \right| X_i \right) \right\rbrace} \right) \right] \geqslant 0 \label{eq:5:ineq_cond_quantile} \\
& \alpha \EE \left[ Y \right] - \EE \left[ Y \ind_{\left\lbrace Y \leqslant q^\alpha \left( Y \right) \right\rbrace} \right] = \EE \left[ \left( Y - q^\alpha \left( Y \right) \right) \left( \alpha-\ind_{\left\lbrace Y \leqslant q^\alpha \left( Y \right) \right\rbrace} \right) \right] \geqslant 0\label{eq:5:ineq_quantile} \\
& \EE \left[ Y \ind_{\left\lbrace Y \leqslant q^\alpha \left( \left. Y \right| X_i \right) \right\rbrace} \right] - \EE \left[ Y \ind_{\left\lbrace Y \leqslant q^\alpha \left( Y \right) \right\rbrace} \right] \geqslant 0 \ . \label{eq:5:num_qosa_index}
\end{align}

This implies that $0 \leqslant S_i^\alpha \leqslant 1$ as already noticed in \citet{fort2016new}. The index $S_i^\alpha$ also has the three following interesting properties.
\begin{prop}\label{prop:5:properties_first_qosa}
\begin{enumerate}
\item[]
\item $S_i^\alpha$ is invariant with respect to translations of the output $Y$.
\item $S_i^\alpha$ is invariant by homothety with strictly positive ratio of the output $Y$.
\item A homothety with strictly negative ratio of the output $Y$ gives the index $S_i^{1-\alpha}$ associated to the $1-\alpha$ level.
\end{enumerate}
\end{prop}
\begin{myproof}[Proof of \Cref{prop:5:properties_first_qosa}]
Let us consider any model $Y=\eta \left( \bX \right)$. \verob{We will denote by $S^{\prime\/\alpha}_i$ the QOSA indices related to a r.v. $Y^\prime$.}
\begin{enumerate}
\item Let $Y' = Y+k, \ k \in \RR$. Then, we have $ q^\alpha \left( Y' \right) =  q^\alpha \left( Y \right) + k$ and  $q^\alpha \left( \left. Y' \right| X_i \right) =  q^\alpha \left( \left. Y \right| X_i \right) + k$. It is easy to check that \verob{$S^{\prime\/\alpha}_i = S_i^\alpha$}.
\item Let $Y' = k \times Y, \ k>0$. Then we have, $ q^\alpha \left( Y' \right) =  k \times q^\alpha \left( Y \right)$ and  $q^\alpha \left( \left. Y' \right| X_i \right) =  k \times q^\alpha \left( \left. Y \right| X_i \right)$. We can easily show that \verob{$S^{\prime\/\alpha}_i = S_i^\alpha$}.
\item Let $Y' = k \times Y, \ k<0$. Then we have, $ q^\alpha \left( Y' \right) =  k \times q^{1-\alpha} \left( Y \right)$ and  $q^\alpha \left( \left. Y' \right| X_i \right) =  k \times q^{1-\alpha} \left( \left. Y \right| X_i \right)$. \verob{It leads to $S^{\prime\/\alpha}_i = S_i^{1-\alpha}$}.
\end{enumerate}
\end{myproof}

Now, we are going to investigate the sum $\cS$ of the first-order QOSA indices:
\[
\cS = \sum_{i=1}^d S_i^\alpha = \dfrac{\displaystyle \sum_{i=1}^d \EE \left[ Y \ind_{\left\lbrace Y \leqslant q^\alpha \left( \left. Y \right| X_i \right) \right\rbrace} \right] - d\EE \left[ Y \ind_{\left\lbrace Y \leqslant q^\alpha \left( Y \right) \right\rbrace} \right]}{\alpha \EE \left[ Y \right] - \EE \left[ Y \ind_{\left\lbrace Y \leqslant q^\alpha \left( Y \right) \right\rbrace} \right]} \ .
\]
We see that $\cS \leqslant 1$ if and only if
\begin{equation}\label{eq:5:borne1}
\sum_{i=1}^d \EE \left[ Y \ind_{\left\lbrace Y \leqslant q^\alpha \left( \left. Y \right| X_i \right) \right\rbrace} \right] - d \EE \left[ Y \ind_{\left\lbrace Y \leqslant q^\alpha \left( Y \right) \right\rbrace} \right] \leqslant \left( \alpha \EE \left[ Y \right] - \EE \left[ Y \ind_{\left\lbrace Y \leqslant q^\alpha \left( Y \right) \right\rbrace} \right] \right) \ .
\end{equation}
Or equivalently:
\verob{\[
\alpha \EE\left[ Y \right] + (d-1) \EE \left[ Y \ind_{\left\lbrace Y \leqslant  q^\alpha \left( Y \right) \right\rbrace} \right] - \sum_{i=1}^d \EE \left[ Y \ind_{\left\lbrace Y \leqslant q^\alpha \left( \left. Y \right| X_i \right) \right\rbrace} \right] \geqslant 0 \ .
\]
}
As proved in the following proposition, $\cS$ is smaller than 1 in the case of an additive model with independent inputs. Unfortunately, this result is not true in the general case as showed with a counterexample in Subsection \ref{subsubsec:5:exponential_product}.

\begin{prop}\label{prop:5:sum_lower_than_1}
Let $\bX = \left( X_1, \ldots, X_d \right)$ \verob{with independent $X_i$'s. Let $Y = m_0 + \sum\limits_{i=1}^d m_i \left( X_i \right)$ be an additive model. }% with $m_i, \ i =1, \ldots, d$, the one-dimensional nonparametric functions operating on each element of the vector $\bX$. 
Then, the sum of the first-order QOSA indices $\cS$ satisfies $\cS \leqslant 1$.
\end{prop}
\begin{myproof}[Proof of \Cref{prop:5:sum_lower_than_1}]
Given a random variable $X$, we denote by $q^\alpha(X)$ its $\alpha$-quantile. For any $i=1, \ldots, d$, let $Xs_{(-i)} = \sum\limits_{\substack{1 \leqslant j \leqslant d \\ j \neq i}} m_j (X_j)$, \verob{ thanks to the independence of the $X_i$'s, we have}
\[
q^\alpha \left( \left. Y \right| X_i \right) =  m_0 + m_i (X_i) + q^\alpha \left( Xs_{(-i)} \right), \textnormal{ and } \{Y \leqslant q^\alpha \left( \left. Y \right| X_i \right) \} = \{Xs_{(-i)} \leqslant q^\alpha \left( Xs_{(-i)} \right) \} \ .
\]

\verob{We have%
\begin{align*}
g(\alpha) &:= \alpha \EE \left[ Y \right] + (d-1) \EE \left[ Y \ind_{\left\lbrace Y \leqslant q^\alpha \left( Y \right) \right\rbrace} \right] - \sum_{i=1}^d \EE \left[ Y \ind_{\left\lbrace Y \leqslant q^\alpha \left( \left. Y \right| X_i \right) \right\rbrace} \right] \\
&= \alpha \EE \left[ Y \right] + (d-1) \EE \left[ Y \ind_{\left\lbrace Y \leqslant q^\alpha \left( Y \right) \right\rbrace} \right] - \sum_{i=1}^d \left( \alpha \times m_0 + \EE \left[ m_i (X_i) \ind_{\left\lbrace Xs_{-(i)} \leqslant q^\alpha \left( Xs_{(-i)} \right) \right\rbrace} \right] \right. \\
&\quad \left. + \ \EE \left[ Xs_{(-i)} \ind_{\left\lbrace Xs_{-(i)} \leqslant q^\alpha \left( Xs_{(-i)} \right) \right\rbrace} \right] \right) \ .
\end{align*}
}
\verob{The independence} of the $X_i$'s implies that $\EE \left[ m_i(X_i) \ind_{\left\lbrace Xs_{(-i)} \leqslant q^\alpha \left( Xs_{(-i)} \right) \right\rbrace} \right] = \alpha \EE \left[ m_i(X_i) \right]$ and thus,
\[
g(\alpha) =  (d-1) \EE \left[ \left( \sum_{j=1}^d m_j (X_j) \right) \ind_{\left\lbrace Y \leqslant q^\alpha \left( Y \right) \right\rbrace} \right] - \sum_{i=1}^d \EE \left[ Xs_{(-i)} \ind_{\left\lbrace Xs_{(-i)} \leqslant q^\alpha \left( Xs_{(-i)} \right)\right\rbrace} \right] \ .
\]
Now, we use Lemma \ref{lem:5:like_TVaR} which gives:
\begin{align*}
(d-1) \EE \left[ \left( \sum_{j=1}^d m_j (X_j) \right) \ind_{\left\lbrace Y \leqslant q^\alpha \left( Y \right) \right\rbrace} \right] &= \sum_{i=1}^{d-1} \left( \EE \left[ m_i(X_i) \ind_{\left\lbrace Y \leqslant q^\alpha \left( Y \right) \right\rbrace} \right] + \EE \left[ Xs_{(-i)} \ind_{\left\lbrace Y \leqslant q^\alpha \left( Y \right) \right\rbrace} \right] \right) \\
&\geqslant \sum_{i=1}^{d-1} \left( \EE \left[ m_i(X_i) \ind_{\left\lbrace Y \leqslant q^\alpha \left( Y \right) \right\rbrace} \right] \right. \\
&\qquad \left. + \ \EE \left[ Xs_{(-i)} \ind_{\left\lbrace Xs_{(-i)}\leqslant q^\alpha \left( Xs_{(-i)} \right) \right\rbrace} \right] \right) \ .
\end{align*}
As a consequence,
\begin{align*}
g(\alpha) &\geqslant \sum_{i=1}^{d-1} \EE \left[ m_i(X_i) \ind_{\left\lbrace Y \leqslant q^\alpha \left( Y \right) \right\rbrace} \right] - \EE \left[ Xs_{(-d)} \ind_{\left\lbrace Xs_{(-d)}\leqslant q^\alpha \left( Xs_{(-d)} \right) \right\rbrace} \right] \\
&= \EE \left[ Xs_{(-d)} \ind_{\left\lbrace Y \leqslant q^\alpha \left( Y \right) \right\rbrace} \right] - \EE \left[ Xs_{(-d)} \ind_{\left\lbrace Xs_{(-d)} \leqslant q^\alpha \left( Xs_{(-d)} \right)\right\rbrace} \right] \\
&\geqslant 0 \textnormal{ using again Lemma \ref{lem:5:like_TVaR}}.
\end{align*}
\end{myproof}
First-order QOSA indices capture only the main effect of the $i$-th input. Following \citet{kala2019quantile}, we consider below higher order QOSA indices. 

\subsection{Higher order QOSA indices}
\citet{kala2019quantile} introduced  second-order QOSA indices in order to assess the impact of the interaction effect of two inputs on the $\alpha$-quantile. He proposes to measure the joint effect of the pair $\left( X_i, X_j \right)$ by:
\[
S_{ij}^\alpha = \dfrac{\displaystyle \min_{\theta \in \RR} \EE \left[ \psi_\alpha \left( Y,\theta \right) \right] - \EE \left[ \min_{\theta \in \RR} \EE \left[ \left. \psi_\alpha \left( Y,\theta \right)\right| X_i, X_j \right] \right]}{\displaystyle \min_{\theta \in \RR} \EE \left[ \psi_\alpha \left(Y,\theta \right) \right]} - S_i^\alpha - S_j^\alpha \ .
\]
Higher-order QOSA indices can be expressed analogously. Hence, one obtain a variance-like decomposition for quantiles in the case of independent inputs:
\begin{equation}\label{eq:5:QOSA_decomposition}
\sum_{i=1}^d S_i^\alpha + \sum_{1 \leqslant i < j \leqslant d} S_{ij}^\alpha + \dots + S_{1, \dots, d}^\alpha = 1 \ .
\end{equation}
Although this decomposition appears similar to that of Sobol indices, it is quite different since it does not stem from an univocal decomposition. %is a key point to underline. Sobol indices stem from the univocal definition of the functional ANOVA decomposition whereas the decomposition above related to the QOSA indices is obtained by construction.

The indices $S_i^\alpha, S_{ij}^\alpha$ and higher order allow to assess thoroughly the impact of each input over the $\alpha$-quantile of the output distribution. However, as for Sobol indices, in the case of a large number of inputs, it would require the evaluation of $2^d - 1$ indices, which could be computationally demanding. Therefore, it is suitable to introduce the so-called total QOSA index as suggested by \citet{kala2019quantile} that measures the contribution of an input, including its main effect as well as its interactions effects, of any order, with other input variables:
\[
ST_i^\alpha = \dfrac{\displaystyle \EE \left[ \min_{\theta \in \RR} \EE \left[ \left. \psi_\alpha \left( Y,\theta \right) \right| \bX_{-i} \right] \right]}{\displaystyle \min_{\theta \in \RR} \EE \left[ \psi_\alpha \left( Y,\theta \right) \right]} = \dfrac{\EE \left[ \psi_\alpha \left( Y, q^\alpha \left( \left. Y \right| \bX_{-i} \right) \right) \right]}{\EE \left[ \psi_\alpha \left( Y, q^\alpha \left( Y \right) \right) \right]} \ .
\]
The total QOSA index may be rewritten as follows
\[
ST_i^\alpha = \dfrac{\alpha \EE \left[ Y \right] - \EE \left[ Y \ind_{\left\lbrace Y \leqslant q^\alpha \left( \left. Y \right| \bX_{-i} \right) \right\rbrace} \right]}{\alpha \EE \left[ Y \right] - \EE \left[ Y \ind_{\left\lbrace Y \leqslant q^\alpha \left( Y \right) \right\rbrace } \right]} \ .
\]
As for the first-order QOSA index, the total one has the three following interesting properties.
\begin{prop}\label{prop:5:properties_total_qosa}
\begin{enumerate}
\item[]
\item $ST_i^\alpha$ is invariant with respect to translations of the output $Y$.
\item $ST_i^\alpha$ is invariant by homothety with strictly positive ratio of the output $Y$.
\item A homothety with strictly negative ratio of the output $Y$ gives the index $ST_i^{1-\alpha}$ associated to the $1-\alpha$ level.
\end{enumerate}
\end{prop}
\begin{myproof}[Proof of \Cref{prop:5:properties_total_qosa}]
Just adapting the steps of the proof of the \Cref{prop:5:properties_first_qosa} for the total QOSA index.
\end{myproof}

\Cref{prop:5:total_upper_first_qosa} below shows that the total QOSA index is greater than or equal to the first-order one for any $\alpha$-level in the case of an additive model with $\bX$ that has independent marginals. This is a major difference from variance-based methods, specifically Sobol indices. Indeed, it is well-known that for a purely additive model with independent inputs, we have for the Sobol indices $ST_i = S_i, \ \forall i \in \cD$, which is not the case for the QOSA indices. It therefore appears that the total QOSA index captures some interaction between the inputs when using an additive model. The origin of this \verob{phenomenom} is not yet understood at this stage and requires further analysis. \\
Besides, it should be noted that \Cref{prop:5:total_upper_first_qosa} is not verified in the general non additive case as emphasized with a counterexample in Subsection \ref{subsubsec:5:lognormal_qosa}.

%\textcolor{Scor_color}{Montrer STi plus grand que Si: discussion avec Véronique. On a que STi est supérieur à Si dans le cas additif. Essayer de voir dans la démo ci qui explique cela. De plus, dans le cas gaussien, pourquoi l'indice d'ordre 1 est constant a été bien compris. Essayer de faire quelque chose pour l'indice total. Trouver un lien comme pour l'ordre 1 avec le quantile.}

\begin{prop}\label{prop:5:total_upper_first_qosa}
Let $\bX = \left( X_1, \ldots, X_d \right)$ with the $X_i$'s independent. Let $Y = m_0 + \sum\limits_{i=1}^d m_i \left( X_i \right)$ be an additive model with $m_i, \ i =1, \ldots, d$, the one-dimensional nonparametric functions operating on each element of the vector $\bX$. Then, 
\[
\forall \ \alpha \in \left] 0,1 \right[, \quad S_i^\alpha \leqslant ST_i^\alpha \ .
\]
\end{prop}
\begin{myproof}[Proof of \Cref{prop:5:total_upper_first_qosa}]
We have:
\[
ST_i^\alpha - S_i^\alpha  = \dfrac{\left( \alpha \EE \left[ Y \right] - \EE \left[ Y \ind_{\left\lbrace Y \leqslant q^\alpha \left( \left. Y \right| \bX_{-i} \right) \right\rbrace} \right] \right) - \left( \EE \left[ Y \ind_{\left\lbrace Y \leqslant q^\alpha \left( \left. Y \right| X_i \right) \right\rbrace} \right] - \EE \left[ Y \ind_{\left\lbrace Y \leqslant q^\alpha \left( Y \right) \right\rbrace} \right] \right)}{\alpha \EE \left[ Y \right] - \EE \left[ Y \ind_{\left\lbrace Y \leqslant q^\alpha \left( Y \right) \right\rbrace } \right]} \ .
\]
As the denominator is \verob{non negative}  according to Equation \eqref{eq:5:ineq_quantile}, we just have to show that the numerator is \verob{also non negative. 

Let}
\[
g(\alpha) = \alpha \EE \left[ Y \right] - \EE \left[ Y \ind_{\left\lbrace Y \leqslant q^\alpha \left( \left. Y \right| \bX_{-i} \right) \right\rbrace} \right]  + \EE \left[ Y \ind_{\left\lbrace Y \leqslant q^\alpha \left( Y \right) \right\rbrace} \right] - \EE \left[ Y \ind_{\left\lbrace Y \leqslant q^\alpha \left( \left. Y \right| X_i \right) \right\rbrace} \right] \ ,
\]
\verob{For any $i=1, \ldots, d$, let $Xs_{(-i)} = \sum\limits_{\substack{1 \leqslant j \leqslant d \\ j \neq i}} m_j (X_j)$,  thanks to the independence, we have}
\begin{align*}
& q^\alpha \left( \left. Y \right| X_i \right) =  m_0 + m_i \left( X_i \right) + q^\alpha \left( Xs_{(-i)} \right), \textnormal{ and } \{Y \leqslant q^\alpha \left( \left. Y \right| X_i \right) \} = \{Xs_{(-i)} \leqslant q^\alpha \left( Xs_{(-i)} \right) \} \ , \\
& q^\alpha \left( \left. Y \right| \bX_{-i} \right) =  m_0 + Xs_{(-i)} + q^\alpha \left( m_i \left( X_i \right) \right), \textnormal{ and } \{Y \leqslant q^\alpha \left( \left. Y \right| \bX_{-i} \right) \} = \{ m_i \left( X_i \right) \leqslant q^\alpha \left( m_i \left( X_i \right) \right) \} \ .
\end{align*}

\verob{Then,} 
\begin{align*}
g(\alpha) &= \alpha \EE \left[ Y \right] - \EE \left[ Y \ind_{\left\lbrace Y \leqslant q^\alpha \left( \left. Y \right| \bX_{-i} \right) \right\rbrace} \right] + \EE \left[ Y \ind_{\left\lbrace Y \leqslant q^\alpha \left( Y \right) \right\rbrace} \right] - \EE \left[ Y \ind_{\left\lbrace Y \leqslant q^\alpha \left( \left. Y \right| X_i \right) \right\rbrace} \right] \\
&= \alpha \EE \left[ Y \right] - \alpha m_0 - \EE \left[ m_i \left( X_i \right) \ind_{\left\lbrace m_i \left( X_i \right) \leqslant q^\alpha \left( m_i \left( X_i \right) \right) \right\rbrace} \right] - \EE \left[ Xs_{(-i)} \ind_{\left\lbrace m_i \left( X_i \right) \leqslant q^\alpha \left( m_i \left( X_i \right) \right) \right\rbrace} \right] \\
&\quad + \EE \left[ Y \ind_{\left\lbrace Y \leqslant q^\alpha \left( Y \right) \right\rbrace} \right] - \alpha m_0 - \EE \left[ Xs_{(-i)} \ind_{\left\lbrace Xs_{(-i)} \leqslant q^\alpha \left( Xs_{(-i)} \right)\right\rbrace} \right] \\
&\quad - \EE \left[ m_i \left( X_i \right) \ind_{\left\lbrace Xs_{(-i)} \leqslant q^\alpha \left( Xs_{(-i)} \right)\right\rbrace} \right] \ .
\end{align*}
Now, the independence of the $X_i$'s implies that $\EE \left[ Xs_{(-i)} \ind_{\left\lbrace m_i \left( X_i \right) \leqslant q^\alpha \left( m_i \left( X_i \right) \right) \right\rbrace} \right] = \alpha \EE \left[ Xs_{(-i)} \right]$ and $\EE \left[ m_i(X_i) \ind_{\left\lbrace Xs_{(-i)} \leqslant q^\alpha \left( Xs_{(-i)} \right) \right\rbrace} \right] = \alpha \EE \left[ m_i(X_i) \right]$. Thus,
\begin{align*}
g(\alpha) &= \EE \left[ \left( \sum_{j=1}^d m_j \left( X_j \right) \right) \ind_{\left\lbrace Y \leqslant q^\alpha \left( Y \right) \right\rbrace} \right] - \EE \left[ m_i \left( X_i \right) \ind_{\left\lbrace m_i \left( X_i \right) \leqslant q^\alpha \left( m_i \left( X_i \right) \right)\right\rbrace} \right] \\
&\quad - \EE \left[ Xs_{(-i)} \ind_{\left\lbrace Xs_{(-i)} \leqslant q^\alpha \left( Xs_{(-i)} \right)\right\rbrace} \right] \\
&= \left( \EE \left[ m_i \left( X_i \right) \ind_{\left\lbrace Y \leqslant q^\alpha \left( Y \right) \right\rbrace} \right] - \EE \left[ m_i \left( X_i \right) \ind_{\left\lbrace m_i \left( X_i \right) \leqslant q^\alpha \left( m_i \left( X_i \right) \right) \right\rbrace} \right] \right) \\
&\quad + \left( \EE \left[ Xs_{(-i)} \ind_{\left\lbrace Y \leqslant q^\alpha \left( Y \right) \right\rbrace} \right] - \EE \left[ Xs_{(-i)} \ind_{\left\lbrace Xs_{(-i)} \leqslant q^\alpha \left( Xs_{(-i)} \right)\right\rbrace} \right] \right) \ .
\end{align*}
Now, the two last terms are positive according to Lemma \ref{lem:5:like_TVaR} which concludes the proof.
\end{myproof}
Several calculations below will help to better understand the behaviour, the usefulness and the limitations of QOSA indices. 

\section{QOSA calculations on special cases}\label{sec:qosa_calculs}

In this section, we compute for some distributions the first-order and total QOSA indices for which we obtain the expressions in a closed or nearly closed-form. In particular, the examples with Gaussian inputs allow to investigate the behavior of the indices when \verob{ there is some} statistical dependence between the inputs.

\subsection{Product of two exponential input variables}\label{subsubsec:5:exponential_product}

Let $Y = X_1 \cdot X_2$, with $X_1 \sim \cE ( \lambda )$, $X_2 \sim \cE (\delta)$, these two variables being independent. \verob{After simple calculations}, we get the first-order QOSA indices
\begin{equation}
S_1^\alpha = S_2^\alpha = 1 - \dfrac{\left(\alpha - 1 \right) \log \left( 1 - \alpha \right)}{\alpha - \lambda \delta \cdot \EE \left[ Y \ind_{\left\lbrace Y \leqslant q^\alpha \left( Y \right) \right\rbrace} \right]} \ ,
\end{equation}
and the total QOSA indices
\begin{equation}
ST_1^\alpha = ST_2^\alpha = \dfrac{\left(\alpha - 1 \right) \log \left( 1 - \alpha \right)}{\alpha - \lambda \delta \cdot \EE \left[ Y \ind_{\left\lbrace Y \leqslant q^\alpha \left( Y \right) \right\rbrace} \right]} \ .
\end{equation}
The term $\EE \left[ Y \ind_{\left\lbrace Y \leqslant q^\alpha \left( Y \right) \right\rbrace} \right]$ can be approximated by using a Monte-Carlo estimation or a numerical integration.

The equality of the first-order and total QOSA indices for both inputs is a particular case due to the exponential distribution. Indeed, let $X_2 \overset{\cL}{=} \frac{\lambda}{\delta} X_1'$ with $X_1'$ an independent copy of $X_1$. Then, the model writes
\begin{align*}
Y & \overset{\cL}{=} \frac{\lambda}{\delta} X_1 \cdot X_1' \\
& \overset{\cL}{=} k Z
\end{align*}
with $Z = X_1 \cdot X_1'$ and $k=\frac{\lambda}{\delta} > 0$. As the inputs $X_1$ and $X_1'$ have the same distribution, their impact over the $\alpha$-quantile of the model output $Z$ is identical. Therefore, by using item 2. of \Cref{prop:5:properties_first_qosa,prop:5:properties_total_qosa} (because $Y$ is just a homothety with strictly positive ratio of $Z$), that \verob{explains} why both first-order and total QOSA indices of the inputs are equal.

Figure \ref{fig:5:expo_prod_2d_first_total_qosa} below presents the behavior of the indices as a function of the level $\alpha$ for the model computed with $\lambda = 1/10$ and $\delta=1$. The truncated expectation is estimated with a Monte-Carlo algorithm and a sample of size $n=10^9$. We observe that the first-order and total QOSA indices vary in opposite directions. The first-order QOSA indices go to 1 when $\alpha$ tends to 1 while the total ones go to 0 when $\alpha$ tends to 1. It is interesting to notice that from $\alpha \approx 0.96$ the total QOSA indices are lower than the first-order ones and the sum of the first-order ones is greater than 1. That corroborates that \Cref{prop:5:properties_first_qosa,prop:5:properties_total_qosa} are not verified outside the additive model context.
\begin{figure}[h]
	\centering
	\makebox[\textwidth][c]{\includegraphics[scale=0.5]{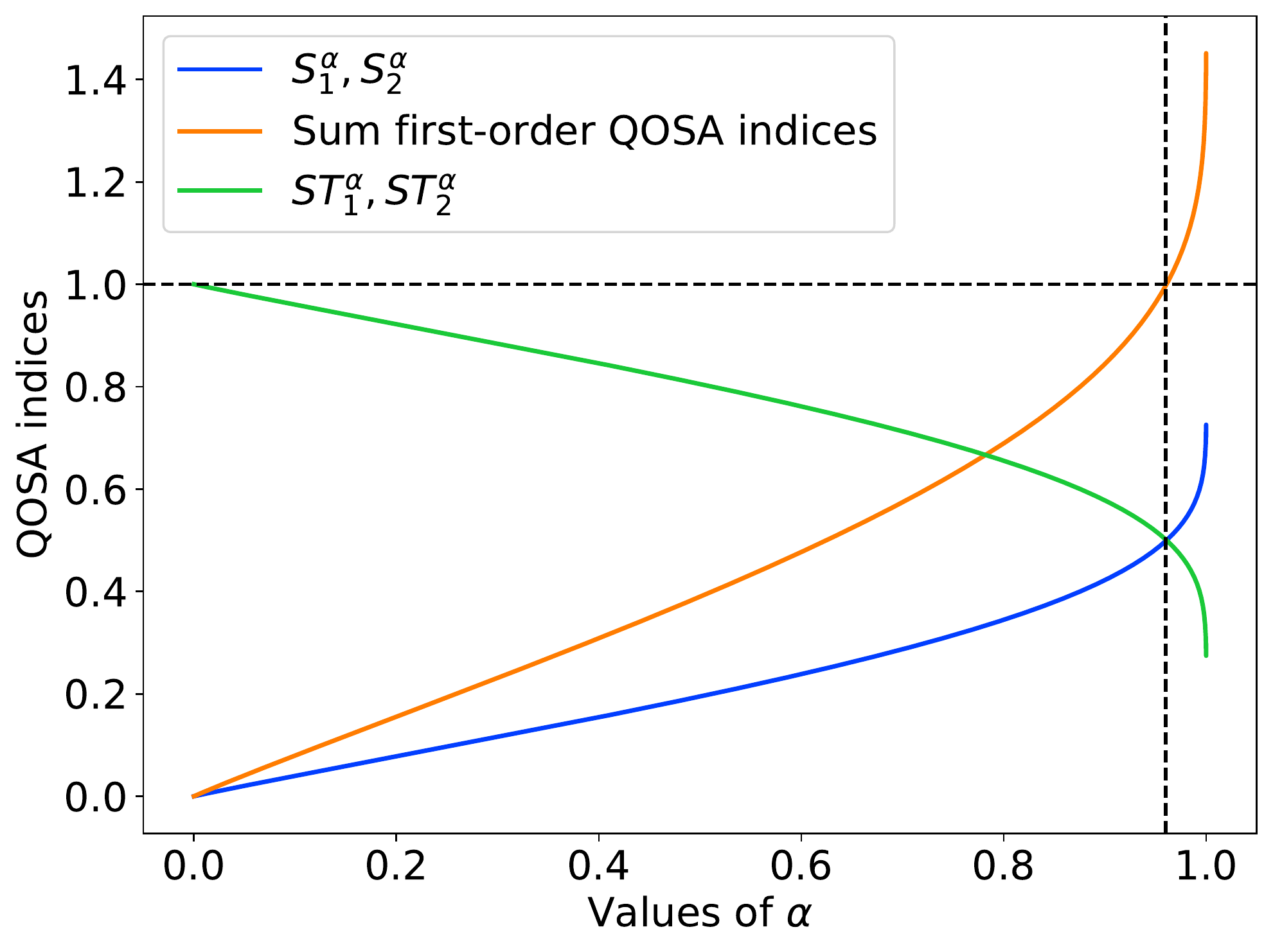}}
	\caption{Evolution of the first-order and total QOSA indices at different levels $\alpha$ for the product of two exponentials with $\lambda = 1/10$ for the fisrt input and $\delta=1$ for the second one.}
	\label{fig:5:expo_prod_2d_first_total_qosa}
\end{figure}

\subsection{Linear model with Gaussian input variables}\label{subsubsec:5:gaussian_qosa}

We study in this subsection a linear model with Gaussian inputs which implies that the resulting output is also Gaussian. This framework facilitates calculations to obtain the analytical values given below.

\begin{prop}
Let $Y=\eta\left( \bX \right) =  \beta_0 + \bbeta^\sfT \bX$ with $\beta_0 \in \RR$, $\bbeta \in \RR^d$ and $\bX \sim \cN(\bmu, \bSigma)$ where $\bSigma \in \RR^{d \times d}$ is a positive-definite matrix, then the first-order and total QOSA indices for the variable $i$ at the $\alpha$-level are
\begin{align}
S_i^\alpha &= 1 - \dfrac{\sqrt{\bbeta_{-i}^\sfT \left( \bSigma_{-i,-i} - \bSigma_{-i,i} \bSigma_{i,i}^{-1} \bSigma_{i,-i} \right) \bbeta_{-i}}}{\sigma_Y} \ , \label{eq:5:first_order_qosa_gaussian} \\
ST_i^\alpha &= \dfrac{\left| \bbeta_{i} \right| \sqrt{\bSigma_{i,i} - \bSigma_{i,-i} \bSigma_{-i,-i}^{-1} \bSigma_{-i,i} }}{\sigma_Y} \ , \label{eq:5:total_order_qosa_gaussian}
\end{align}
with $\sigma_Y^2 = \var \left( Y \right) = \bbeta^\sfT \bSigma \bbeta$.
\end{prop}

We observe that as $\beta_0$ and $\bmu$ are translation parameters, they do not have any influence. Nevertheless, no general conclusion can be drawn from \Cref{eq:5:first_order_qosa_gaussian,eq:5:total_order_qosa_gaussian} except that the values of the first-order and total QOSA indices are the same for all levels $\alpha$. This phenomenon is specific to the Gaussian linear model and will be detailed hereafter in dimension $2$.  
Indeed, a look at the case $d=2$ may help to understand why  QOSA indices does not depend on $\alpha$-level in the Gaussian framework.
%We shall consider%
\vero{This feature will be analyzed by studying} only the impact of the variable $X_1$.

We have that
\[
Y | X_1 \sim \cN \left( \beta_1 X_1 + \beta_2 \EE \left[ \left. X_2 \right| X_1 \right], \ \beta_2^2 \var \left( \left. X_2 \right| X_1 \right) \right) \ .
\] 
As we work with Gaussian distributions, the conditional variance $\var \left( \left. X_2 \right| X_1 \right)$ does not depend on the specific value of $X_1$ and is $\var \left( \left. X_2 \right| X_1 \right) = \sigma_2^2 \left( 1 - \rho^2 \right)$. The conditional quantile of $Y$ given $X_1$ has the following expression
\[ 
q^\alpha \left( \left. Y \right| X_1 \right) =  \beta_1 X_1 + \beta_2 \EE \left[ \left. X_2 \right| X_1 \right] + \left| \beta_2 \right| \sqrt{\var \left( \left. X_2 \right| X_1 \right)} \Phi^{-1} \left( \alpha \right)
\]
\verob{with $\Phi$ the standard normal distribution function. One way to assess the impact of the variable $X_1$ on the quantile $q^\alpha \left( Y \right)$ would be to calculate the ratio
%
%\[
%\dfrac{q^\alpha \left( \left. Y \right| X_1 \right) - \left( \beta_1 X_1 + \beta_2 \EE \left[ \left. X_2 \right| X_1 \right] \right)}{q^\alpha \left( Y \right) - \EE \left[ Y \right]} \ ,
%\]
%
%for several values of $x_1$ and to observe its evolution. If, when one sets $X_1$ to several different values, the ratio varies a lot, then $X_1$ is highly responsible for the value $q^\alpha \left( Y \right)$. The issue with this measure is the choice of the value $x_1$ of $X_1$, which can be solved by considering the expectation of this quantity
%
\[
\dfrac{\EE \left[ q^\alpha \left( \left. Y \right| X_1 \right) \right] - \EE \left[ Y \right]}{q^\alpha \left( Y \right) - \EE \left[ Y \right]} \ .
\]
}
Thus, by using that $q^\alpha \left( Y \right) =  \EE \left[ Y \right] + \sigma_Y \Phi^{-1} \left( \alpha \right)$, the previous ratio equals
\[
\dfrac{\EE \left[ q^\alpha \left( \left. Y \right| X_1 \right) \right] - \EE \left[ Y \right]}{q^\alpha \left( Y \right) - \EE \left[ Y \right]} = \dfrac{\left| \beta_2 \right| \sigma_2 \sqrt{\left( 1 - \rho^2 \right)}}{\sigma_Y} \ .
\]
%
%As $Y$ is a Gaussian distribution as well as the conditional distribution of $Y$ given $X_1$, then the influence of the input $X_1$ over the quantile $q^\alpha \left( Y \right)$ is only driven by the variances as showed in \eqref{eq:5:cst_quantile_gaussien}. We therefore recognize the term involved in the first-order QOSA index $S_1^\alpha$ in \eqref{eq:5:first_order_qosa_gaussian_2d} as well as in the total QOSA index $ST_2^\alpha$ in \eqref{eq:5:total_order_qosa_gaussian_2d}.
A simple calculation shows that $ S_1^\alpha =\displaystyle 1 - \dfrac{\left| \beta_2 \right| \sigma_2 \sqrt{1 - \rho^2}}{\sigma_Y}$ and gives the relation with QOSA indices. In a more general way, the following equality holds for all variables $i=1, \ldots, d$ when using a \textbf{linear Gaussian model} and explains why the first-order and total QOSA indices do not depend on the $\alpha$-level:
\[
\dfrac{\alpha \EE \left[ Y \right] - \EE \left[ Y \ind_{\left\lbrace Y \leqslant q^\alpha \left( \left. Y \right| X_i \right) \right\rbrace} \right]}{\alpha \EE \left[ Y \right] - \EE \left[ Y \ind_{\left\lbrace Y \leqslant q^\alpha \left( Y \right) \right\rbrace} \right]} = \dfrac{\EE \left[ q^\alpha \left( \left. Y \right| X_i \right) \right] - \EE \left[ Y \right]}{q^\alpha \left( Y \right) - \EE \left[ Y \right]} \ .
\]

\begin{figure}[h]
	\centering
	\makebox[\textwidth][c]{\includegraphics[width=1.\textwidth]{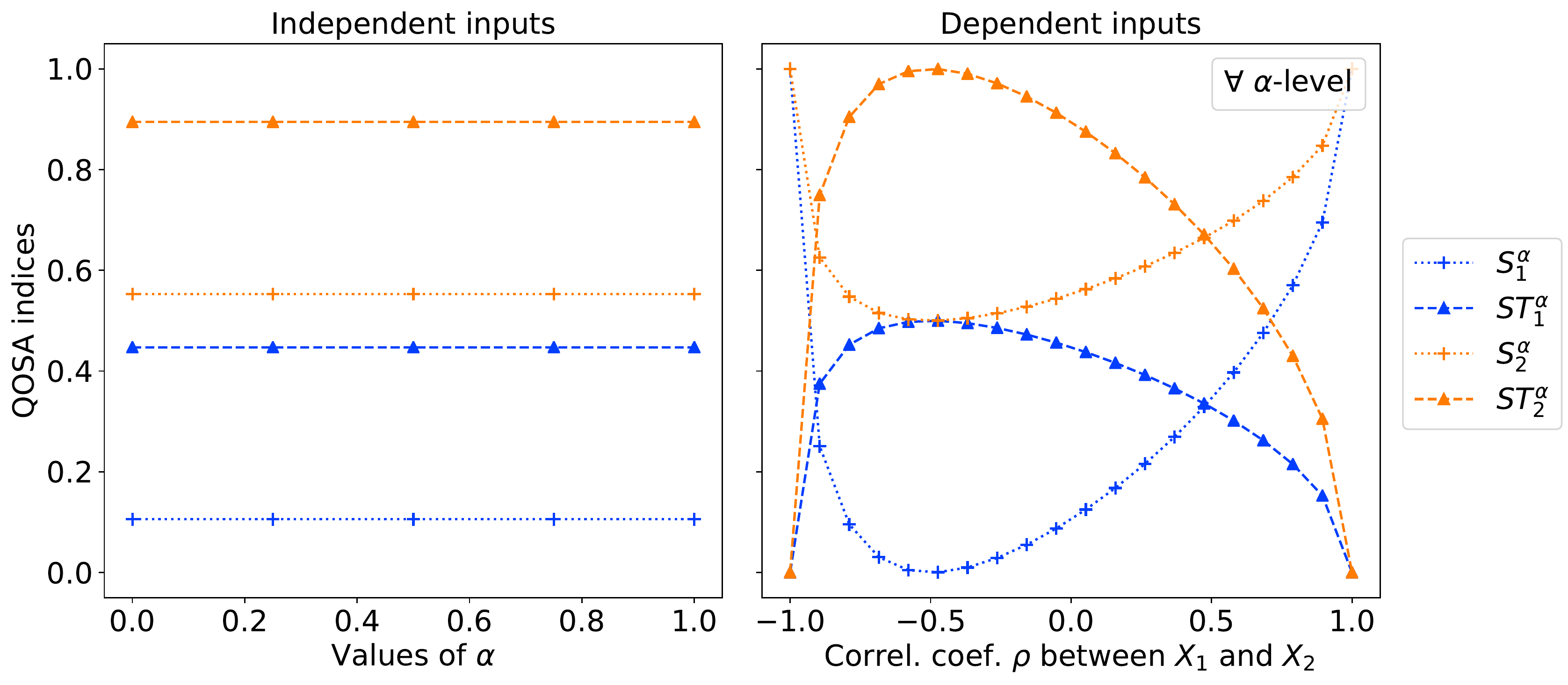}}
	\caption{First-order and total QOSA indices with independent (resp. dependent) inputs on the left (resp. right) graph.}
	\label{fig:5:gaussian_2d_first_total_qosa}
\end{figure}

We now study the particular case $\mu_1=\mu_2=0,\beta_1=\beta_2=1,\sigma_1=1$ and $\sigma_2=2$. The analytical values of the indices are depicted in Figure \ref{fig:5:gaussian_2d_first_total_qosa} on the left-hand graph for independent inputs and on the right-hand plot as a function of the correlation coefficient between the two inputs in order to investigate the influence of the dependence.

For the independent case, it appears that the variable $X_2$ has the higher impact over the $\alpha$-quantile, which is consistent with the setting established. Besides, we have $S_i^\alpha \leqslant ST_i^\alpha, \ i=1,2$ as proved in \Cref{prop:5:total_upper_first_qosa}. \\
Regarding the dependent case, we observe that the total QOSA indices tend to zero as $\left| \rho \right| \rightarrow 1$. It is also interesting \verob{to notice that} $ST_i^\alpha \leqslant S_i^\alpha$ for some correlation coefficients. The behaviour of these indices is similar to that of the Sobol indices in the context of dependent inputs as studied in \citet{kucherenko2012estimation,iooss2019shapley}. Indeed, by making an analogy with the method proposed by \citet{mara2015non} based on four Sobol indices, we could say that in the case of dependent inputs: 
\begin{itemize}
\item the first-order QOSA index describes the influence of a variable including its dependence with other variables,
\item the total QOSA index describes the influence of a variable without its dependence with other variables.
\end{itemize}

\subsection{Gaussian input variables, \verop{Lognormal output}}\label{subsubsec:5:lognormal_qosa}

We analyze in this subsection a model with Gaussian inputs whose output is a Log-normal distribution so that we no longer have identical indices for any $\alpha$-level. Using Gaussian inputs makes calculations possible and we obtain the following analytical values.

\begin{prop}
Let $Y=\eta\left( \bX \right) =  \exp \left( \beta_0 + \bbeta^\sfT \bX \right)$ with $\beta_0 \in \RR$, $\bbeta \in \RR^d$ and $\bX \sim \cN(\bmu, \bSigma)$ where $\bSigma \in \RR^{d \times d}$ is a positive-definite matrix, then the first-order and total QOSA indices for the variable $i$ at the $\alpha$-level are
\begin{align}
S_i^\alpha &= 1 - \dfrac{\alpha - \Phi \left( \Phi^{-1} \left( \alpha \right) - \sqrt{\bbeta_{-i}^\sfT \left( \bSigma_{-i,-i} - \bSigma_{-i,i} \bSigma_{i,i}^{-1} \bSigma_{i,-i} \right) \bbeta_{-i}} \right)}{\alpha - \Phi \left( \Phi^{-1} \left( \alpha \right) - \sigma \right)} \ , \label{eq:5:first_order_qosa_lognormal} \\
ST_i^\alpha &= \dfrac{\alpha - \Phi \left( \Phi^{-1} \left( \alpha \right) - \left| \bbeta_{i} \right| \sqrt{\bSigma_{i,i} - \bSigma_{i,-i} \bSigma_{-i,-i}^{-1} \bSigma_{-i,i}} \right)}{\alpha - \Phi \left( \Phi^{-1} \left( \alpha \right) - \sigma \right)} \ , \label{eq:5:total_order_qosa_lognormal}
\end{align}
with $\sigma^2 = \bbeta^\sfT \bSigma \bbeta$ \verob{and $\Phi$ the  standard normal distribution function}.
\end{prop}

We observe that $\beta_0$ and $\bmu$ do not play any role as these are scale parameters in this example. %While the indices vary as a function of $\alpha$ in this scheme, no conclusion can be reached from \Cref{eq:5:first_order_qosa_lognormal,eq:5:total_order_qosa_lognormal}. As a consequence, l
Let us consider the particular case $d=2$ with
\[
\bmu = 
\begin{pmatrix}
\mu_1 \\
\mu_2 \\
\end{pmatrix},
\ \bbeta = 
\begin{pmatrix}
\beta_1 \\
\beta_2 \\
\end{pmatrix}
\textnormal{ and }
\bSigma = 
\begin{pmatrix}
\sigma_1^2 & \rho \sigma_1 \sigma_2 \\
\rho \sigma_1 \sigma_2 & \sigma_2^2
\end{pmatrix},
\ -1 \leq \rho \leq 1,\ \sigma_1 > 0,\ \sigma_2 > 0.
\]
We have $\sigma^2 = \beta_1^2 \sigma_1^2 + 2 \rho \beta_1 \beta_2 \sigma_1 \sigma_2 + \beta_2^2 \sigma_2^2$ and obtain from \Cref{eq:5:first_order_qosa_lognormal,eq:5:total_order_qosa_lognormal}
\begin{equation}\label{eq:5:first_order_qosa_lognormal_2d}
\begin{split}
S_1^\alpha &= 1 - \dfrac{\alpha - \Phi \left( \Phi^{-1} \left( \alpha \right) - \left| \beta_2 \right| \sigma_2 \sqrt{1 - \rho^2} \right)}{\alpha - \Phi \left( \Phi^{-1} \left( \alpha \right) - \sigma \right)} \ , \\
S_2^\alpha &= 1 - \dfrac{\alpha - \Phi \left( \Phi^{-1} \left( \alpha \right) - \left| \beta_1 \right| \sigma_1 \sqrt{1 - \rho^2} \right)}{\alpha - \Phi \left( \Phi^{-1} \left( \alpha \right) - \sigma \right)} \ ,
\end{split}
\end{equation}
and
\begin{equation}\label{eq:5:total_order_qosa_lognormal_2d}
\begin{split}
ST_1^\alpha &= \dfrac{\alpha - \Phi \left( \Phi^{-1} \left( \alpha \right) - \left| \beta_1 \right| \sigma_1 \sqrt{1 - \rho^2} \right)}{\alpha - \Phi \left( \Phi^{-1} \left( \alpha \right) - \sigma \right)} \ , \\
ST_2^\alpha &= \dfrac{\alpha - \Phi \left( \Phi^{-1} \left( \alpha \right) -\left| \beta_2 \right| \sigma_2 \sqrt{1 - \rho^2} \right)}{\alpha - \Phi \left( \Phi^{-1} \left( \alpha \right) - \sigma \right)} \ .
\end{split}
\end{equation}

\begin{figure}[h!]
	\centering
	\makebox[\textwidth][c]{\includegraphics[width=1.\textwidth]{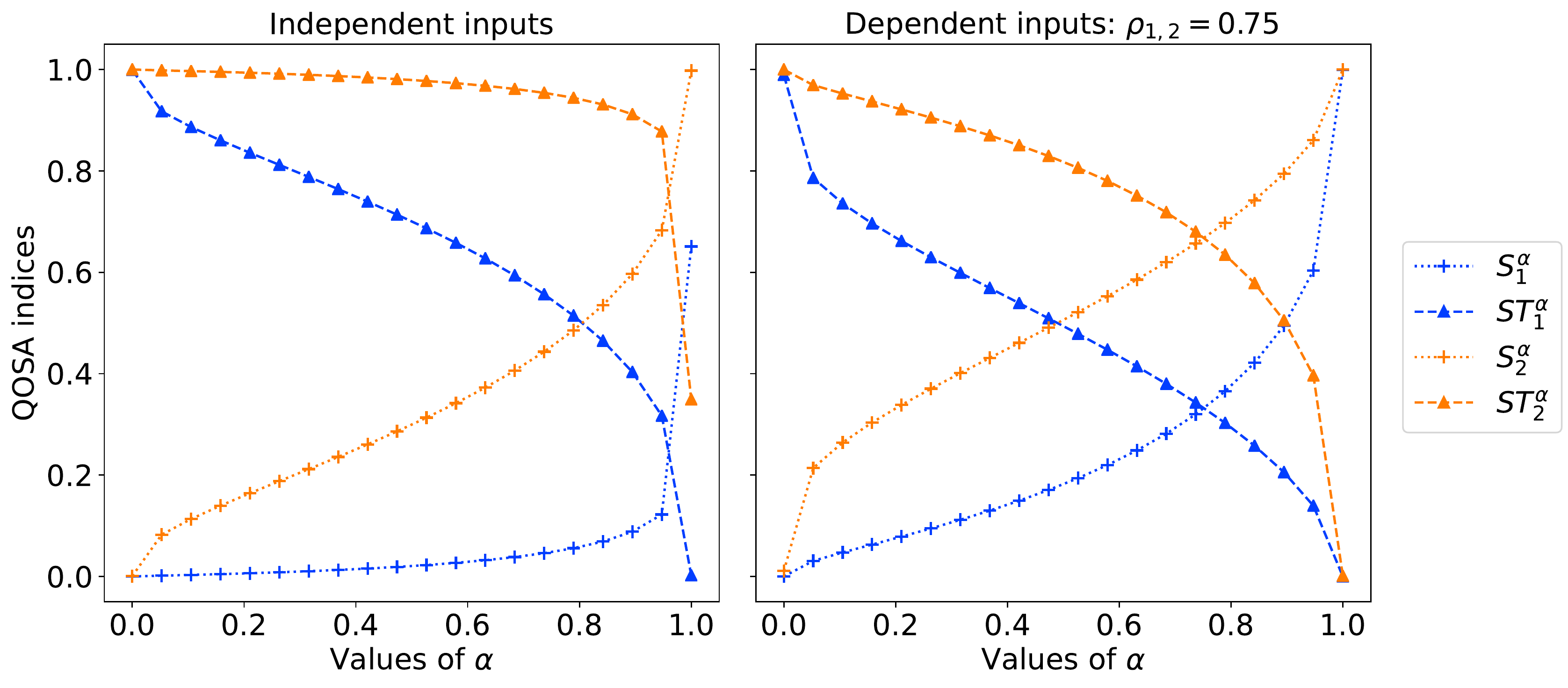}}
	\caption{First-order and total QOSA indices with independent (resp. dependent) inputs on the left (resp. right) graph.}
	\label{fig:5:lognormal_2d_first_total_qosa}
\end{figure}

In all further tests, we take $\mu_1=\mu_2=0,\beta_1=\beta_2=1,\sigma_1=1$ and $\sigma_2=2$. \\
Figure \ref{fig:5:lognormal_2d_first_total_qosa} presents the analytical values of the first-order and total QOSA indices for both independent inputs and correlated inputs with $\rho_{1,2} = 0.75$. In the independent setting, the influence of the variable $X_1$ is close to 0 except for large \verob{values of $\alpha$}. We also note that the first-order and total QOSA indices vary in reverse direction and from some $\alpha$-level, $ST_i^\alpha \leqslant S_i^\alpha, \ i=1,2$. This supports that \Cref{prop:5:total_upper_first_qosa} is not true outside the additive framework with independent inputs. \\
The behavior of the indices is similar in the dependent case. However, the influence of the input $X_1$ is reinforced in this scheme due to its large correlation with $X_2$ that is an influent variable. Indeed, the index $S_1^\alpha$ increases faster than in independent case. On the contrary, the index $ST_2^\alpha$ decreases to 0 quicker than in the independent case because of its high dependence with $X_1$.

\begin{figure}[h]
	\centering
	\makebox[\textwidth][c]{\includegraphics[width=1.\textwidth]{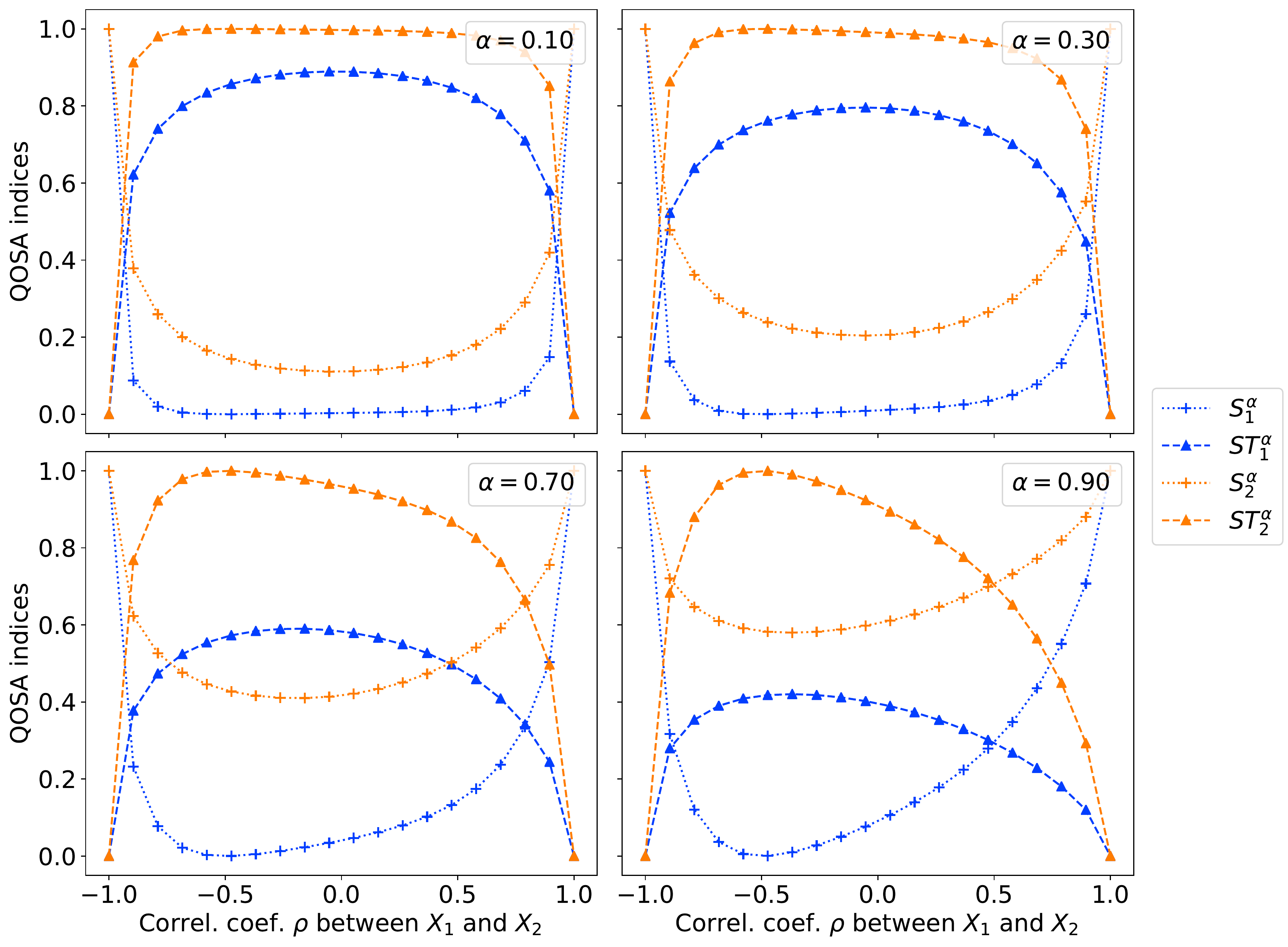}}
	\caption{Evolution of the first-order and total QOSA indices at different values of $\rho$ for several levels $\alpha$.}
	\label{fig:5:lognormal_2d_first_total_qosa_correlation}
\end{figure}

To get another perspective on the impact of the dependence over the indices, we plot in Figure \ref{fig:5:lognormal_2d_first_total_qosa_correlation}, for several levels $\alpha$, the evolution of the latter as a function of the correlation coefficient. As for the linear Gaussian model, we observe that the total QOSA indices tend to zero as $\left| \rho \right| \rightarrow 1$ and they are lower than the first-order ones for some correlation coefficients.

%\textcolor{Scor_color}{Ajouter comme dans le papier de Kucherenko (Estimation of global sensitivity indices for models with dependent variables) un graphe montrant l'évolution de l'indice en fonction du coefficient de corrélation pour un modèle tridimensionnel en prenant la même matrice de corrélation. En conclusion, on espère observer comme pour les Sobol que l'indice total tend vers 0 quand la valeur absolue de $\rho$ tend vers 1. Et aussi des cas avec les indices totaux inférieurs aux indices d'ordre 1.}

Hence, we have $S_i \leqslant ST_i^\alpha, \ i=1,\ldots,d$ for additive models with independent inputs. But this context is far from reality for many concrete examples and this inequality is no longer valid outside this framework as outlined by examples presented in Subsections \ref{subsubsec:5:exponential_product} and \ref{subsubsec:5:lognormal_qosa}. This therefore makes the interpretation of the indices complicated. \\
Furthermore, in the case of dependent inputs, the behaviour of the QOSA indices should be compared to that \verob{of Sobol indices}. Indeed, whatever the model (additive or not), it may happen in this scheme that the first-order QOSA indices are higher than the total ones depending on \verob{the correlation level}. We have also observed that total indices tend to zero as the absolute value of the correlation goes to 1.% These similar phenomenons observed for the Sobol indices were elucidated by \citet{mara2015non} thanks to a method based on the calculation of four sensitivity indices per input (see Subsection \ref{subsec:3:sobol_indices_dt}). 

We could establish a strategy similar to \citet{mara2015non} in order to better understand the impact of inputs in case of statistical dependence over the $\alpha$-quantile, i.e., if their contribution derives from their marginal importance or their dependence with another variable. But, we prefer to turn to  Shapley values  which present good properties for both independent and dependent inputs. Indeed, they allocate fairly to each input the interaction and/or dependency effect in which it is involved.

\section{Quantile oriented Shapley effects}\label{sec:6:shapley_qosa}

In this section, we propose to use  Shapley values defined in Equation \eqref{eq:3:shapley_value}, \verob{ and recalled below,} in order to quantify the impact of each input over the $\alpha$-quantile of the output distribution
\begin{equation}\label{eq:6:shapley_value}
v^{i} = \sum_{\cJ \subseteq \cD \backslash \{i\}} \dfrac{(d - |\cJ| - 1)!|\cJ|!}{d!} \left( c \left( \cJ \cup \{i\} \right) - c \left( \cJ \right) \right) \ ,
\end{equation}
with $c( \cdot )$ a generic cost function which maps the exploratory power generated by each subset $\cJ \subseteq \cD$.

Shapley value was first adapted within the framework of variance-based sensitivity measures to measure how much of $\var \left( Y \right)$ can be attributed to each $X_i$. Indeed, \citet{owen2014sobol} and \citet{song2016shapley} proposed to use the two following unnormalized cost functions to measure the variance of $Y$ caused by the uncertainty of the inputs in the subset $\cJ \subseteq \cD$ also named as being the explanatory power created by $\cJ$:
\begin{equation}\label{eq:6:cost_functions}
\tilde{c}(\cJ) = \var \left( \EE \left[ Y | \bX_{\cJ} \right] \right) \textnormal{ and } c(\cJ) = \EE \left[ \var \left( Y | \bX_{-\cJ} \right) \right] \ .
\end{equation}
Measuring the variance of $Y$ caused by the uncertainty of the inputs in $\cJ$ is equivalent to assess the impact of the inputs over the expected output. Thus, when using the cost functions given in \eqref{eq:6:cost_functions}, the feature of interest of the output considered is the expectation denoted by $\theta^* \left( Y \right) = \EE \left[ Y \right]$. We show in the left-hand column in \Cref{tab:6:cost_function_shapley} that both cost functions may be rewritten according to the contrast function related to the expectation as well as the conditional feature $\theta^* \left( \left. Y \right| X_\cJ \right) = \EE \left[ \left. Y \right| \bX_{\cJ} \right]$ for the first cost function and $\theta^* \left( \left. Y \right| X_{-\cJ} \right) = \EE \left[ \left. Y \right| \bX_{-\cJ} \right]$ for the second one.

\begin{table}[h]
\centering
\rowcolors{1}{lightgray}{white}
\begin{tabular}{C{6.5 cm} I C{6.5 cm}}
\multicolumn{2}{c}{\textbf{Feature of interest}} \\
$\theta^* \left( Y \right) = \EE \left[ Y \right]$ & $\theta^* \left( Y \right) = q^\alpha \left( Y \right) $ \\
\multicolumn{2}{c}{\textbf{Contrast function}} \\
$\psi \left( y, \theta \right) = \left( y - \theta \right)^2$ & $\psi \left( y, \theta \right) = \left( y - \theta \right) \left( \alpha - \ind_{\left\lbrace y \leqslant \theta \right\rbrace} \right)$ \\
\multicolumn{2}{c}{\textbf{Average contrast function}} \\
$\begin{aligned}
\var \left( Y \right) &= \EE \left[ \psi \left(Y, \ \EE \left[ Y \right] \right) \right] \\
					  &= \EE \left[ \psi \left(Y, \ \theta^* \left( Y \right) \right) \right]
\end{aligned}$ & $\begin{aligned} 
\Upsilon \left( Y \right) &=  \EE \left[ \psi \left(Y, \ q^\alpha \left( Y \right) \right) \right]\\
						  &=\EE \left[ \psi \left(Y, \ \theta^* \left( Y \right) \right) \right] 
\end{aligned}$ \\
\multicolumn{2}{c}{\textbf{First cost function}} \\
$\begin{aligned}
\tilde{c}(\cJ) &= \var \left( \EE \left[ \left. Y \right| \bX_{\cJ} \right] \right) \\
			   &= \EE \left[ \psi \left( \EE \left[ \left. Y \right| \bX_{\cJ} \right], \ \EE \left[ Y \right] \right) \right] \\
			   &= \EE \left[ \psi \left( \theta^* \left( \left. Y \right| X_\cJ \right), \ \theta^* \left( Y \right) \right) \right] \\
			   & 
\end{aligned}$
$\begin{aligned}
\tilde{c}(\emptyset) = 0 \qquad \tilde{c}(\cD) = \var \left( Y \right)
\end{aligned}$ & $\begin{aligned}
\tilde{c}(\cJ) &= \EE \left[ \psi \left( \theta^* \left( \left. Y \right| X_\cJ \right), \ \theta^* \left( Y \right) \right) \right] \\
			   &= \EE \left[ \psi \left( q^\alpha \left( \left. Y \right| \bX_{\cJ} \right), \ q^\alpha \left( Y \right) \right) \right] \\
			   & \\
			   &
\end{aligned}$
$\begin{aligned}
\tilde{c}(\emptyset) = 0 \qquad \tilde{c}(\cD) = \Upsilon \left( Y \right)
\end{aligned}$ \\
\multicolumn{2}{c}{\textbf{Second cost function}} \\
$\begin{aligned}
c(\cJ) &= \EE \left[ \var \left( \left. Y \right| \bX_{-\cJ} \right) \right] \\
	   &= \EE \left[ \psi \left( Y, \ \EE \left[ \left. Y \right| \bX_{-\cJ} \right] \right) \right] \\
	   &= \EE \left[ \psi \left( Y, \ \theta^* \left( \left. Y \right| X_{-\cJ} \right) \right) \right] \\
	   &
\end{aligned}$
$\begin{aligned}
c(\emptyset) = 0 \qquad c(\cD) = \var \left( Y \right)
\end{aligned}$ & $\begin{aligned} 
c(\cJ) &= \EE \left[ \psi \left( Y, \ \theta^* \left( \left. Y \right| X_{-\cJ} \right) \right) \right] \\
	   &= \EE \left[ \psi \left( Y, \ q^\alpha \left( \left. Y \right| \bX_{-\cJ} \right) \right) \right] \\
       & \\
	   &
\end{aligned}$
$\begin{aligned}
c(\emptyset) = 0 \qquad c(\cD) = \Upsilon \left( Y \right)
\end{aligned}$ \\
\hline
\end{tabular}
\captionof{table}{Analogy of the cost functions used for quantifying the impact of the inputs over the expectation for the case where the quantile is the feature of interest.} 
\label{tab:6:cost_function_shapley}
\end{table}

Thus, in order to define indices for another feature of interest, the idea is to substitute the contrast function of the expectation by that associated with the feature of interest required.

Let us now use $\theta^* \left( Y \right),\ \theta^* \left( \left. Y \right| X_\cJ \right)$ and $\theta^* \left( \left. Y \right| X_{-\cJ} \right)$ for $\cJ \subseteq \cD$ as generic expressions to designate a feature of interest and the conditional ones related to a contrast function $\psi$. The impact of the inputs over $\theta^* \left( Y \right)$ is therefore assessed by measuring their contribution to the averaged contrast function $\EE \left[ \psi \left(Y, \ \theta^* \left( Y \right) \right) \right]$. This one can be seen as a relevant distance allowing to quantify the variability around the feature of interest. The contributions of the inputs are then calculated with the following cost functions measuring the explanatory power of the subset $\cJ \subseteq \cD$
\begin{equation}\label{eq:6:generic_cost_functions}
\tilde{c}(\cJ) = \EE \left[ \psi \left( \theta^* \left( \left. Y \right| X_\cJ \right), \ \theta^* \left( Y \right) \right) \right] \textnormal{ and } c(\cJ) = \EE \left[ \psi \left( Y, \ \theta^* \left( \left. Y \right| X_{-\cJ} \right) \right) \right] \ .
\end{equation}
These cost functions are valid choices if they satisfy that the empty set creates no value, and that all inputs generate $\EE \left[ \psi \left(Y, \ \theta^* \left( Y \right) \right) \right]$. This is, \vero{for example, verified for all contrast functions listed in \citet{fort2016new} which allow to propose new indices named \textit{\textbf{Goal oriented Shapley effects (GOSE)}}.} \\
\vero{In our context, this property is verified in particular for cost functions related to quantiles presented in \Cref{tab:6:cost_function_shapley}.} Hence we may propose \textit{\textbf{Shapley effects subordinated to quantiles}}.
%Following this strategy, we propose to quantify the impact of the inputs over the $\alpha$-quantile of the model output by using the corresponding contrast function as well as the two unnormalized cost functions established in the right-hand column in \Cref{tab:6:cost_function_shapley}. 
However, we define the \textit{\textbf{Quantile oriented Shapley effects (QOSE)}} denoted by $Sh_i^\alpha$ \verob{with  the second cost function} because it verifies that the incremental cost $c \left( \cJ \cup \{i\} \right) - c \left( \cJ \right)$ is non negative, \verop{which is also a desirable property for cost functions}. Indeed, for $\cJ \subseteq \cD \backslash \{i\}$, we have
\begin{align*}
c \left( \cJ \cup i \right) - c \left( \cJ \right) &= \left( \alpha \EE \left[ Y \right] - \EE \left[ Y \ind_{\left\lbrace Y \leqslant q^\alpha \left( \left. Y \right| \bX_{-\cJ \cup i} \right) \right\rbrace} \right] \right) - \left( \alpha \EE \left[ Y \right] - \EE \left[ Y \ind_{\left\lbrace Y \leqslant q^\alpha \left( \left. Y \right| \bX_{-\cJ} \right) \right\rbrace} \right] \right) \\
&= \EE \left[ \left( Y - q^\alpha \left( \left. Y \right| \bX_{- \cJ \cup i} \right) \right) \left( \ind_{\left\lbrace Y \leqslant q^\alpha \left( \left. Y \right| X_{-\cJ} \right) \right\rbrace} - \ind_{\left\lbrace Y \leqslant q^\alpha \left( \left. Y \right| \bX_{- \cJ \cup i} \right) \right\rbrace} \right) \right] \geqslant 0 \ .
\end{align*}
At this stage, this property has not yet been demonstrated for the first cost function $\tilde{c}$.

We study in the next subsections examples whose analytical values of the index $Sh_i^\alpha$ are computed by using the cost function normalized by the quantity $\Upsilon \left( Y \right)$ introduced in \Cref{tab:6:cost_function_shapley}, so that $\displaystyle \sum_{i=1}^d Sh_i^\alpha =1$. Our aim is to show that these new indices give sensible answers compared to the classical QOSA indices defined in Section \ref{sec:5:qosa_indices}.

\subsection{Linear model with Gaussian input variables}

We obtain the following analytical values for the linear model with Gaussian inputs.
\begin{prop}
If $Y=\eta\left( \bX \right) =  \beta_0 + \bbeta^\sfT \bX$ with $\beta_0 \in \RR$, $\bbeta \in \RR^d$ and $\bX \sim \cN(\bmu, \bSigma)$ where $\bSigma \in \RR^{d \times d}$ is a positive-definite matrix, then the Quantile oriented Shapley effect for the variable $i$ at the $\alpha$-level is
\begin{equation}\label{eq:6:shapley_qosa_gaussian}
\begin{split}
Sh_i^\alpha &= \dfrac{1}{d \cdot \sigma_Y} \sum_{\cJ \subseteq \cD \backslash \{i\}} \binom{d-1}{\left| \cJ \right|}^{-1} \left[ \sqrt{\bbeta_{\cJ + i}^\sfT \left( \bSigma_{\cJ + i, \cJ + i} - \bSigma_{\cJ + i,-\cJ - i} \bSigma_{-\cJ - i,-\cJ - i}^{-1} \bSigma_{-\cJ - i,\cJ + i} \right) \bbeta_{\cJ + i}} \right. \\
&\quad \left. - \sqrt{\bbeta_{\cJ}^\sfT \left( \bSigma_{\cJ, \cJ} - \bSigma_{\cJ,-\cJ} \bSigma_{-\cJ,-\cJ}^{-1} \bSigma_{-\cJ,\cJ} \right) \bbeta_{\cJ}}  \right]
\end{split}
\end{equation}
with $\sigma_Y^2 = \var \left( Y \right) = \bbeta^\sfT \bSigma \bbeta$, and $\cJ + i$ (resp. $-\cJ -i$), a notational compression for $\cJ \cup \{i\}$ (resp. $-\cJ \cup \{i\}$).
\end{prop}

As for the QOSA index, we may notice that $\beta_0$ and $\bmu$ do not play any role as translation parameters and that the index does not depend on the $\alpha$-level which is a specificity of the linear Gaussian model as explained previously. Let us consider the case $d=2$ with
\[
\bmu = 
\begin{pmatrix}
\mu_1 \\
\mu_2 \\
\end{pmatrix},
\ \bbeta = 
\begin{pmatrix}
\beta_1 \\
\beta_2 \\
\end{pmatrix}
\textnormal{ and }
\bSigma = 
\begin{pmatrix}
\sigma_1^2 & \rho \sigma_1 \sigma_2 \\
\rho \sigma_1 \sigma_2 & \sigma_2^2
\end{pmatrix},
\ -1 \leq \rho \leq 1,\ \sigma_1 > 0,\ \sigma_2 > 0.
\]

We have $\sigma_Y^2 = Var \left( Y \right) = \beta_1^2 \sigma_1^2 + 2 \rho \beta_1 \beta_2 \sigma_1 \sigma_2 + \beta_2^2 \sigma_2^2$ and obtain from\eqref{eq:6:shapley_qosa_gaussian}
\begin{equation}\label{eq:6:shapley_qosa_gaussian_2d}
\begin{split}
Sh_1^\alpha &= \dfrac{1}{2} - \dfrac{\left| \beta_2 \right| \sigma_2 \sqrt{1 - \rho^2}}{2 \cdot \sigma_Y} + \dfrac{\left| \beta_1 \right| \sigma_1 \sqrt{1 - \rho^2}}{2 \cdot \sigma_Y} \ , \\
Sh_2^\alpha &= \dfrac{1}{2} - \dfrac{\left| \beta_1 \right| \sigma_1 \sqrt{1 - \rho^2}}{2 \cdot \sigma_Y} + \dfrac{\left| \beta_2 \right| \sigma_2 \sqrt{1 - \rho^2}}{2 \cdot \sigma_Y} \ .
\end{split}
\end{equation}
We observe that the correlation effects on the first-order QOSA indices (e.g. $ \sigma_Y - \left| \beta_2 \right| \sigma_2 \sqrt{1 - \rho^2}$ for $X_1$) and on the total QOSA indices (e.g. $\left| \beta_1 \right| \sigma_1 \sqrt{1 - \rho^2}$ for $X_1$) are allocated half to the Quantile oriented Shapley effects - QOSE. We also see that the Shapley effects are equal when the correlation is maximum (i.e. $\left| \rho \right| = 1 $). 

\begin{figure}[h!]
	\centering
	\makebox[\textwidth][c]{\includegraphics[width=1.\textwidth]{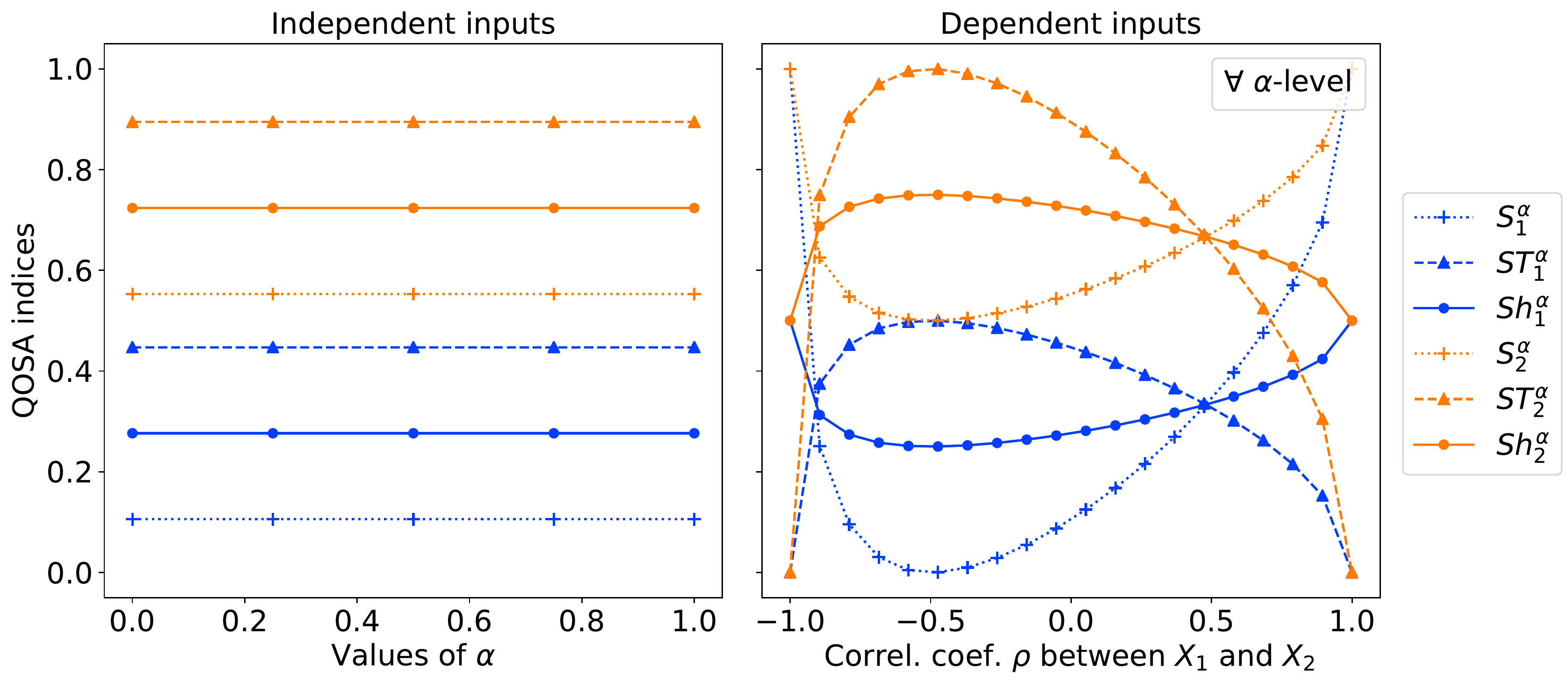}}
	\caption{First-order and total QOSA indices as well as the QOSE with independent (resp. dependent) inputs on the left (resp. right) graph.}
	\label{fig:6:gaussian_2d_first_total_shapley_qosa}
\end{figure}

Figure \ref{fig:6:gaussian_2d_first_total_shapley_qosa} presents the first-order and total QOSA indices \verob{as well as the QOSE} for the particular case $\mu_1=\mu_2=0, \ \beta_1=\beta_2=1, \ \sigma_1=1$ and $\sigma_2=2$. \\
\verob{On} \verop{the left-hand} \verob{graph of the figure, we see that the Shapley effects are also constant and they are brackected by the first-order and total QOSA indices : $S_i^\alpha \leqslant Sh_i^\alpha \leqslant ST_i^\alpha, \ i=1,2$.}%This is due to the fact that the first-order QOSA index omits interaction effects, while the total one overcounts them relative to Shapley that fairly shares this effect on each variable involved within.

We illustrate on the right-hand graph the evolution of the indices as a function of the correlation between the two inputs. As $X_2$ is the more uncertain variable, its sensitivity indices are larger than those of $X_1$. Then, although the values are not identical, we can note that the shape of the curves is exactly the same as that observed for the variance-based Shapley effects calculated for the two-dimensional Gaussian linear model (with the same setting) in \citet{iooss2019shapley}. Indeed, we observe that in the presence of correlation, \verop{the QOSE} lie between the first-order QOSA indices and the total ones with either $S_i^\alpha \leqslant Sh_i^\alpha \leqslant ST_i^\alpha$ or $ST_i^\alpha \leqslant Sh_i^\alpha \leqslant S_i^\alpha, \ i=1,2$. This phenomenon is called the ``sandwich effect'' within the variance framework in \citet{iooss2019shapley}. Finally, as for the variance-based Shapley effects, it also seems that the dependence between the two inputs lead to a rebalancing of \verob{their respective QOSE}.

%\begin{align*}
%QSh_1^\alpha = \dfrac{1}{2} \left( S_1^\alpha + ST_1^\alpha \right) \\
%QSh_2^\alpha = \dfrac{1}{2} \left( S_2^\alpha + ST_2^\alpha \right) \\
%\end{align*}

\subsection{Gaussian input variables, \verob{Lognormal output}}

We analyze in this subsection the analytical values below for the model with Gaussian inputs and the resulting output Log-normal distributed.
\begin{prop}
If $Y=\eta\left( \bX \right) =  \exp \left( \beta_0 + \bbeta^\sfT \bX \right)$ with $\beta_0 \in \RR$, $\bbeta \in \RR^d$ and $\bX \sim \cN(\bmu, \bSigma)$ where $\bSigma \in \RR^{d \times d}$ is a positive-definite matrix, then the Quantile oriented Shapley effect for the variable $i$ at the $\alpha$-level is
\begin{equation}\label{eq:6:shapley_qosa_lognormal}
\begin{split}
Sh_i^\alpha &= \dfrac{1}{d \cdot A} \sum_{\cJ \subseteq \cD \backslash \{i\}} \binom{d-1}{\left| \cJ \right|}^{-1} \left[ \Phi \left( \Phi^{-1} \left( \alpha \right) - B \left( \cJ \right) \right) - \Phi \left( \Phi^{-1} \left( \alpha \right) - C \left( \cJ, i\right) \right) \right]
\end{split}
\end{equation}
with
\begin{align*}
A &= \alpha - \Phi \left( \Phi^{-1} \left( \alpha \right) - \sigma \right) \textnormal{ and } \sigma^2 = \bbeta^\sfT \bSigma \bbeta \ , \\
B \left( \cJ \right) &= \sqrt{\bbeta_{\cJ}^\sfT \left( \bSigma_{\cJ, \cJ} - \bSigma_{\cJ,-\cJ} \bSigma_{-\cJ,-\cJ}^{-1} \bSigma_{-\cJ,\cJ} \right) \bbeta_{\cJ}} \ , \\
C \left( \cJ, i \right) &= \sqrt{\bbeta_{\cJ + i}^\sfT \left( \bSigma_{\cJ + i, \cJ + i} - \bSigma_{\cJ + i,-\cJ - i} \bSigma_{-\cJ - i,-\cJ - i}^{-1} \bSigma_{-\cJ - i,\cJ + i} \right) \bbeta_{\cJ + i}} \ ,
\end{align*}
where $\cJ + i$ (resp. $-\cJ -i$) is a notational compression for $\cJ \cup \{i\}$ (resp. $-\cJ \cup \{i\}$).
\end{prop}

As for the QOSA indices, we observe that $\beta_0$ and $\bmu$ do not play any role and that the indices depend on $\alpha$ compared to the linear Gaussian model. However, it is difficult to reach a conclusion from \Cref{eq:6:shapley_qosa_lognormal}. Accordingly, we consider the particular case $d=2$ with
\[
\bmu = 
\begin{pmatrix}
\mu_1 \\
\mu_2 \\
\end{pmatrix},
\ \bbeta = 
\begin{pmatrix}
\beta_1 \\
\beta_2 \\
\end{pmatrix}
\textnormal{ and }
\bSigma = 
\begin{pmatrix}
\sigma_1^2 & \rho \sigma_1 \sigma_2 \\
\rho \sigma_1 \sigma_2 & \sigma_2^2
\end{pmatrix},
\ -1 \leq \rho \leq 1,\ \sigma_1 > 0,\ \sigma_2 > 0.
\]

We have $\sigma^2 = \beta_1^2 \sigma_1^2 + 2 \rho \beta_1 \beta_2 \sigma_1 \sigma_2 + \beta_2^2 \sigma_2^2$ and obtain from \eqref{eq:6:shapley_qosa_lognormal}
\begin{equation}\label{eq:6:shapley_qosa_lognormal_2d}
\begin{split}
Sh_1^\alpha &= \dfrac{1}{2} + \dfrac{1}{2} \cdot \dfrac{\Phi \left( \Phi^{-1} \left( \alpha \right) - \left| \beta_2 \right| \sigma_2 \sqrt{1 - \rho^2} \right)}{\alpha - \Phi \left( \Phi^{-1} \left( \alpha \right) - \sigma \right)} - \dfrac{1}{2} \cdot \dfrac{\Phi \left( \Phi^{-1} \left( \alpha \right) - \left| \beta_1 \right| \sigma_1 \sqrt{1 - \rho^2} \right)}{\alpha - \Phi \left( \Phi^{-1} \left( \alpha \right) - \sigma \right)} \ , \\
Sh_2^\alpha &= \dfrac{1}{2} + \dfrac{1}{2} \cdot \dfrac{\Phi \left( \Phi^{-1} \left( \alpha \right) - \left| \beta_1 \right| \sigma_1 \sqrt{1 - \rho^2} \right)}{\alpha - \Phi \left( \Phi^{-1} \left( \alpha \right) - \sigma \right)} - \dfrac{1}{2} \cdot \dfrac{\Phi \left( \Phi^{-1} \left( \alpha \right) - \left| \beta_2 \right| \sigma_2 \sqrt{1 - \rho^2} \right)}{\alpha - \Phi \left( \Phi^{-1} \left( \alpha \right) - \sigma \right)} \ .
\end{split}
\end{equation}

We adopt the next settings in all further tests: $\mu_1=\mu_2=0,\beta_1=\beta_2=1,\sigma_1=1$ and $\sigma_2=2$.

\begin{figure}[h]
	\centering
	\makebox[\textwidth][c]{\includegraphics[width=1.\textwidth]{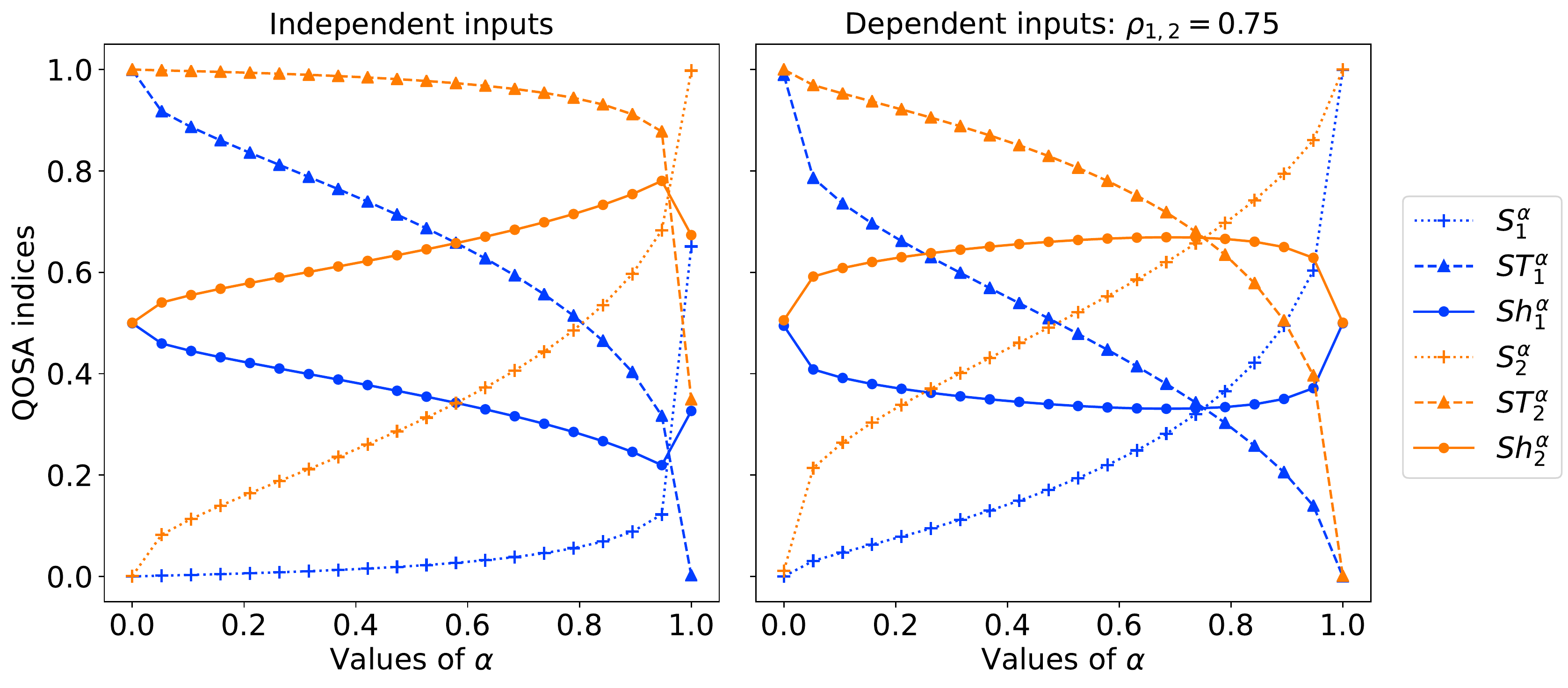}}
	\caption{First-order and total QOSA indices as well as the Quantile oriented Shapley effects with independent (resp. dependent) inputs on the left (resp. right) graph.}
	\label{fig:6:lognormal_2d_first_total_shapley_qosa}
\end{figure}

The analytical values of the first-order, total QOSA indices \verob{and the QOSE} are illustrated in Figure \ref{fig:6:lognormal_2d_first_total_shapley_qosa} for both independent inputs and correlated inputs with $\rho_{1,2} = 0.75$. The ``sandwich effect'' which was noticed in the linear Gaussian model in the presence of correlation is also observed here. Indeed, both in the dependent and independent cases and for all the levels $\alpha$, \verop{the QOSE} lie between the first-order and total QOSA indices.

Besides, with the three indices, we obtain the same ranking of the inputs for all $\alpha$-levels but \verop{the QOSE} is easier to interpret because it properly condenses all the information (dependence and interaction effects). For instance, let us focus over the input $X_1$ on the right-hand graph at the level $\alpha=0.2$. If we use the first-order QOSA index $S_1^\alpha$, we conclude that the impact of the input $X_1$ is low, but not so small because, conversely its total QOSA index is high enough. But, ultimately, it is difficult to quantify precisely on the basis of these two indices the contribution of the input $X_1$ at level $\alpha = 0.2$. The Shapley index, in contrast, contains the marginal contribution of the variable but also those due to dependence and interaction effects that are correctly allocated to it. It therefore makes easier to express an opinion on the impact of the variable by taking into account all possible contributions. This observation is valid for all the levels $\alpha$.

\begin{figure}[h]
	\centering
	\makebox[\textwidth][c]{\includegraphics[width=1.\textwidth]{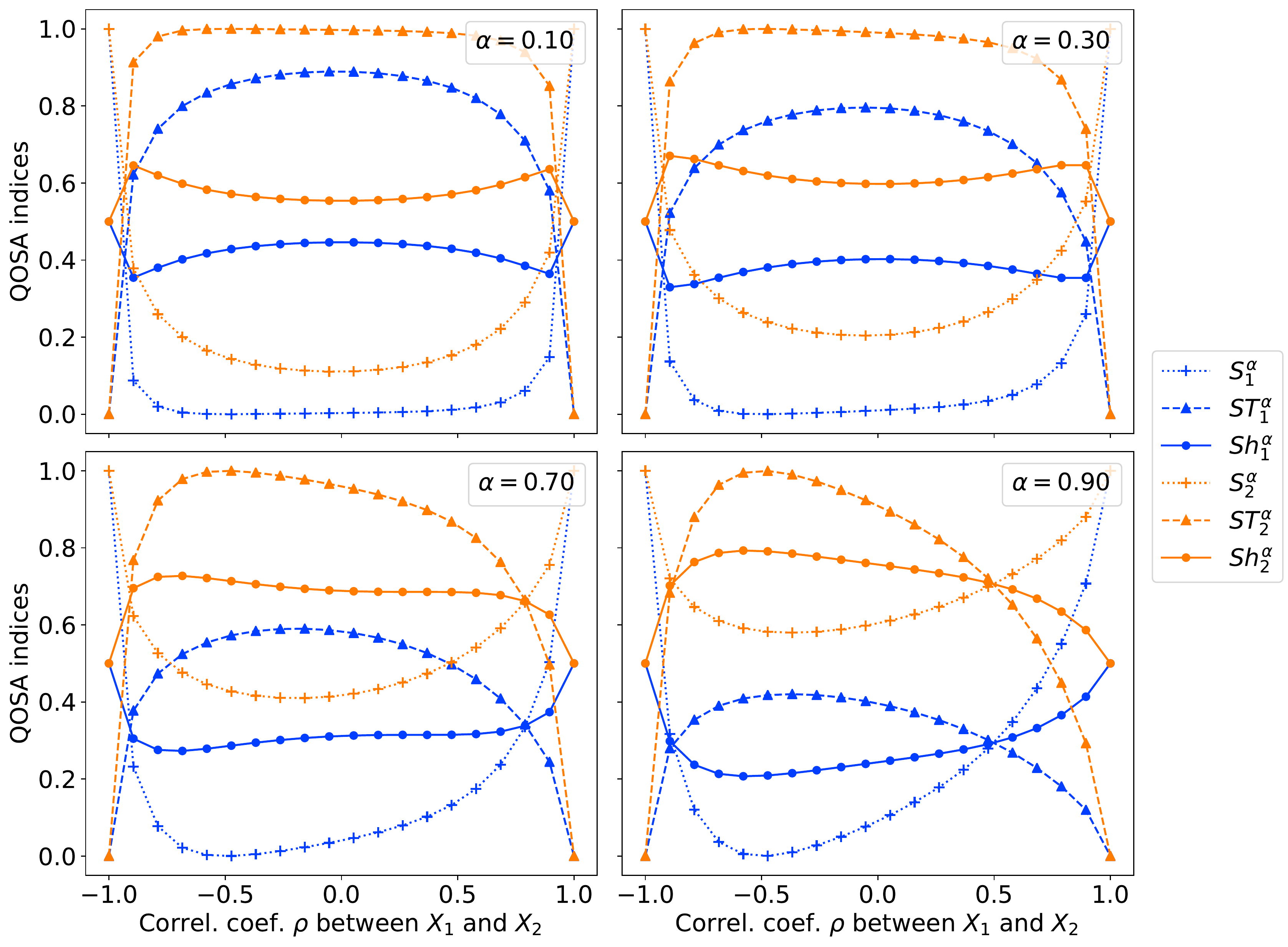}}
	\caption{Evolution of the first-order and total QOSA indices as well as the Quantile oriented Shapley effects at different values of $\rho$ for several levels $\alpha$.}
	\label{fig:6:lognormal_2d_first_total_shapley_qosa_correlation}
\end{figure}

Again, to get another insight  \verob{on the impact of the dependence} over the indices, we plot in Figure \ref{fig:6:lognormal_2d_first_total_shapley_qosa_correlation}, for several levels $\alpha$, the evolution of the latter as a function of the correlation coefficient. As explained before, \verob{the QOSE} give a condensed information of all contributions. That explains why we observe that the Shapley effects of both variables are almost equal for small values of $\alpha$. Conversely, for large values, the variable $X_2$ is the most influential overall except when $\left| \rho \right| \rightarrow 1$ where both inputs have the same contribution.

\section{Perspective}\label{sec:conclusion}
As a conclusion of this preliminary work, QOSE appear to be a good alternative to the classical QOSA indices. Indeed, they make possible to overcome the various problems encountered with the QOSA indices such as $ST_i^\alpha \leqslant S_i^\alpha$ outside the additive framework or when using dependent inputs. Besides, the Sobol indices stem, for example, from the functional ANOVA decomposition but there is no such a decomposition for the quantiles. Hence, as an allocation method, these indices therefore allow to quantify precisely the contribution of each input at the $\alpha$-quantile while taking into account interaction and dependency effects. \\
An additional work would be to \verob{study the QOSE for Gaussian examples in higher dimension. Last but not least, the development of an estimation algorithm to compute these new indices would be a significant contribution.}%, but also a difficult task because they involve all subsets of the inputs.

\clearpage
\bibliographystyle{apalike}
\bibliography{bibliography}

\end{document}